\documentclass[11pt,review,1p,number,sort&compress,a4paper]{elsarticle}
\usepackage{graphicx}
\usepackage{lscape}
\usepackage{longtable,array}
\usepackage{amsmath}
\usepackage{capt-of}
\usepackage{setspace}

\usepackage[dvipsnames]{xcolor}
\graphicspath{{/}}
\DeclareGraphicsExtensions{.pdf,.png,.jpg,}

\newcommand*{\datum}[1]{#1}

\newcommand{\cm}{cm$^{-1}$}
\newcommand{\mc}{\multicolumn}

\newcommand{\Marvel}{{\sc Marvel}}

\newcommand{\A}{A\,$^3\Pi$}
\newcommand{\X}{X\,$^3\Sigma^-$}
\newcommand{\Sa}{a\,$^1\Delta$}
\newcommand{\Sb}{b\,$^1\Sigma^+$}
\newcommand{\Sc}{c\,$^1\Pi$}
\newcommand{\Sd}{d\,$^1\Sigma^+$}

\newcommand{\NoSO}{18}
\newcommand{\NoTR}{3002}

\newcommand{\NoExclTR}{7}
\newcommand{\NoPrCoTR}{2954}
\newcommand{\NoPrCoEL}{1058}
\newcommand{\NoPpEL}{44}
\newcommand{\NoPmEL}{818}
\newcommand{\NoSEL}{17}
\newcommand{\NoUEL}{179}
\newcommand{\NoFlCoTR}{48}
\newcommand{\NoFlCoEL}{62}
\newcommand{\NoX}{542}
\newcommand{\NoA}{403}
\newcommand{\Noc}{58}
\newcommand{\Noa}{55}

\begin{document}

\title{MARVEL analysis of the measured high-resolution spectra of $^{14}$NH} 

\author[ucl]{Daniel Darby-Lewis}
\author[Preston]{Het Shah}
\author[Preston]{Dhyeya Joshi}
\author[Preston]{Fahd Khan} 
\author[Preston]{Miles Kauwo}
\author[Preston]{Nikhil Sethi} 
\author[OldD]{Peter F. Bernath}
\author[Eotvos]{Tibor Furtenbacher}
\author[Eotvos]{Roland T\'obi\'as}
\author[Eotvos]{Attila G. Cs\'asz\'ar}
\author[ucl]{Jonathan Tennyson\corref{cor1}}
\ead{j.tennyson@ucl.ac.uk}

\cortext[cor1]{Jonathan Tennyson}

\address[ucl]{Department of Physics and Astronomy, University College London,
Gower Street, London WC1E 6BT, United Kingdom}
\address[Preston]{Preston Manor School, Carlton Avenue East, Wembley, HA9 8NA, 
United Kingdom}
\address[OldD]{Department of Chemistry and Biochemistry, Old Dominion University, 
4541 Hampton Boulevard, Norfolk, VA 23529, USA}
\address[Eotvos]{Institute  of Chemistry, ELTE E\"otv\"os Lor\'and University 
and MTA-ELTE Complex Chemical Systems Research Group,
H-1518 Budapest 112, P.O. Box 32, Hungary}
\date{\today}

\begin{abstract}
Rovibronic energy levels are determined for four  low-lying electronic states
(\X, \A, \Sa, and \Sc) of the imidogen free radical ($^{14}$NH) using the
\Marvel\  (Measured Active Rotational-Vibrational Energy Levels) technique.  
Compilation of transitions from both laboratory measurements and solar spectra,
found in \NoSO\ publications, yields a dataset of \NoTR\ rovibronic
transitions forming elements of a measured spectroscopic network (SN).
At the end of the MARVEL procedure, the majority of the transitions form a single, 
self-consistent SN component of \NoPrCoTR\  rovibronic transitions and
%with only \NoExclTR\  transitions  excluded. This yields  
\NoPrCoEL\   energy levels,
\NoX, \NoA, and \Noc\  for the \X, \A, and \Sc\  electronic states, respectively.
The \Sa\ electronic state is characterized by \Noa\  $\Lambda$-doublet levels,
counting each level only once. Electronic structure computations show that unusually
the  CCSD(T) method does not accurately predict the \Sa\
excitation energy even at the complete basis set limit.
\label{sec.abst}
 
\end{abstract}

\maketitle
\newpage
\section{Introduction}
\label{sec.intro}

\begin{figure}
\includegraphics[width=0.4\textwidth]{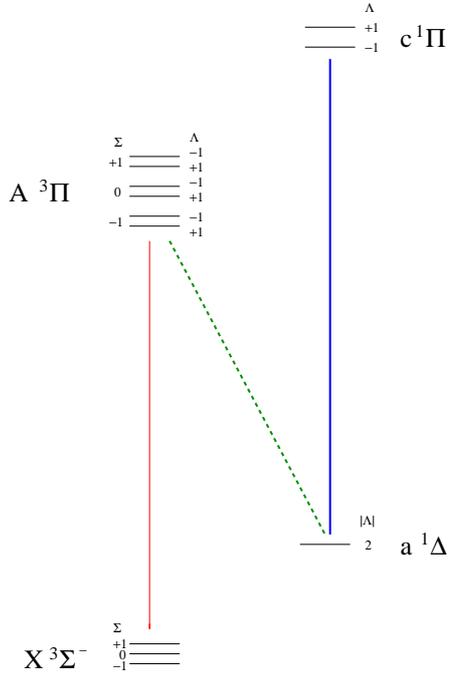}
\caption{Schematic representation of the three electronic band
systems of NH analyzed in this work, with Hund's case $a$ quantum
numbers spin ($\Sigma$) and orbital ($\Lambda$) projections
considered for each state separately; note that these quantum numbers
are in general not good ones for NH.
The total project is given by $\Omega = |\Lambda +\Sigma|$.
The spin-forbidden intercombination band system is depicted using a dashed line.}
\label{fig:NHbands}
\end{figure}

Imidogen, the NH free radical, is easily formed by flash photolysis of ammonia,
often in excited electronic states \cite{75MaGiVe,85MaHiMo}.  
The spectral signature of NH is observed in a variety of plasmas, including
laser-produced plasmas in air \cite{18HaBrPh.NH} and liquids \cite{17PfOuBe.NH},
plasmas used for waste water treatment \cite{18HaLiCh.NH}, and it also serves
to measure ammonia concentration in plasmas \cite{19ZhGaLi.NH}.
Emissions can also be observed   from NH trace species 
due to impurities in or seeding of fusion plasmas \cite{18PeChAk.NH,18PaDiDr.NH}.
NH emissions are also observable in flames \cite{18BrNiNa.NH,19LaGaDe.NH}.
NH has also received considerable attention as an ultracold species that can be 
trapped using buffer gas cooling, magnetic trapping, or Zeeman relaxation
\cite{07CaTsLu.NH,10TsCaHu.NH,11WaLoZu.NH,13JaVaGr.NH} and has been
the subject of detailed, state-resolved energy transfer experiments
\cite{02RiGe.NH,06Kajita.NH}.
Lasing action has also been observed involving particular rovibronic states of
NH \cite{79SmRo}.

NH is an  important astronomical species, whose 
\A--\X\  spectrum was first recorded in the laboratory by Eder in 1893 \cite{93Eder}.
This electronic transition has been detected in the Sun \cite{18FoGr.NH,73GrSa}, in the
comet Cunningham \cite{41SwElBa.NH}, in the interstellar medium by its absorption against 
various stars \cite{72LaBe.NH,91MeRo.NH,97CrWi.NH,09WeGaBe.NH}, 
and in the atmospheres of cool stars themselves \cite{97AoTs.NH}.
The \Sc--\Sa\  band system was originally observed by Dieke and Blue in 1934 \cite{34DiBl}
(see Figure~\ref{fig:NHbands}).
%\blue{AGC: How about A--a (indicated in Figure 1)?}
Line lists are available which provide 
line strengths for rovibrational and rotational transitions within the \X\  ground state of
$^{14}$NH \cite{14BrBeWe.NH,15BrBeWe.NH} and for the \A--\X\  band system \cite{18FeBeHo.NH}. 

In this paper we analyse all the available high-resolution spectra of
NH recorded in the laboratory
\cite{70MaBrGu,75RaLi,76WaRa,78GrLe,82BeAm,82RaSa,82VaMeDy,84UbMeDy,85HaAdKaCu,86RaBe,86BrRaBe,
  86LeEvBr,86UbMeTeDy,86BoBrChGu,87VaDaBrEv,90HaMi,97KlTaWi,99RaBeHi,99RiGe,03VaSaMoJo,04FlBrMaOd,
  04LeBrWiSi,07RoBrFlZi,10RaBe} and obtained from spectroscopic
observations of the Sun \cite{10RaBe,90GrLaSaVa,91GeSaGrFa}.  
The purpose of this work is to provide accurate empirical energy levels
with dependable uncertainties for $^{14}$NH using the Measured Active
Rotational-Vibrational Energy Levels (\Marvel) approach
\cite{jt412,07CsCzFu.marvel,12FuCsi.method,jt750} (though the
traditional name refers to rovibrational states, there is nothing in
the original proposal of \Marvel\ \cite{jt412} which prevents the use
of rovibronic transitions in a \Marvel\ analysis).
\Marvel\ energy levels have been used in the past to compute accurate,
temperature-dependent ideal-gas thermodynamic data \cite{jt661}, to
facilitate the empirical adjustment of potential energy surfaces
\cite{jt743}, and to improve the accuracy of computed line lists
\cite{jt734,19HuScLe.SO2}.

The configuration of the ground electronic state of NH is
$1\sigma^2 2\sigma^2 3\sigma^2 1\pi^2$.
The doubly occupied $\pi$ orbital gives rise to the \X\  ground electronic state
and the low-lying metastable excited \Sa\  and \Sb\  states. 
Rearrangement of the electrons to 
$1\sigma^2 2\sigma^2 3\sigma^1 1\pi^3$ leads to two further low-lying 
electronic states designated \A\ and \Sc. 
{\it Ab initio} calculations show that these are indeed the dominant configurations
of these five electronic states \cite{07OwJaKw,16SoShSu}.
Following some earlier studies \cite{71NeSc,75MeRo},
% CITE: S. V. O'Neil AND H. F. {Schaefer III}, J. Chem. Phys. 55, 394 (1971)
% CITE: W. Meyer AND P. Rosmus, J. Chem. Phys. 63, 2356 (1975)
Owono Owono {\it et al.} \cite{07OwJaKw} provide a comprehensive set of {\it ab initio}
potential energy curves (PEC) for the states considered here. 
The work of Owono Owono {\it et al.} \cite{07OwJaKw} is complemented by a high-accuracy 
study of the  \X\  ground-state PEC by Koput \cite{15Koput} and by other calculations 
\cite{16SoShSu}. 
%, who also considered spin-orbit couplings.

This work started out by concentrating on the five lowest-lying electronic states of NH. 
However, although the metastable \Sb\  state  has been observed 
spectroscopically \cite{75MaGiVe,78GrLe}, there are insufficient data to warrant
a \Marvel\  study including this state's energy levels. 
Analysis of this state therefore awaits the results of further laboratory measurements.
Figure~\ref{fig:NHbands} illustrates the electronic states and the electronic band systems
of NH which we consider in this work. We also revisit the issue of obtaining
accurate electronic exciation energies using standard quantum chemistry methods for
this 8 electron system.

%\newpage
\section{Methodological details}
\subsection{\Marvel}

The \Marvel\ approach \cite{jt412,07CsCzFu.marvel,12FuCsi.method,jt750} enables
a set of assigned experimental transition wavenumbers to be converted
into empirical energy levels with associated uncertainties that are propagated
from the input transitions to the output energy levels.
This conversion relies on the construction of a spectroscopic network
(SN) \cite{11CsFuxx.marvel,12FuCsxx.marvel,14FuArMe.marvel,16ArPeFu.marvel},
which contains all measured and assigned inter-connected transitions.
For a detailed description of the approach, algorithm, and code,
we refer the reader to Ref.~\cite{12FuCsi.method}.

The \Marvel\ approach has been used to validate compiled experimental rovibronic
transitions for the astronomically important species $^{48}$Ti$^{16}$O \cite{jt672},
which proved to be an important step in providing a greatly improved
line list for TiO \cite{jt760}.
Although the spectroscopy of TiO is somewhat more complicated, there are similarities 
between this study and the one presented in Ref.~\cite{jt672}.
Other \Marvel\ studies on chemically or astronomically important molecules include
those for $^{12}$C$_2$ \cite{jt637}, $^{12}$C$_2$H$_2$ \cite{jt705}, $^{14}$NH$_3$
\cite{jt608,jtNH3update}, three isotopologues of SO$_2$ \cite{jt704}, 
H$_2$$^{32}$S \cite{jt718}, $^{90}$Zr$^{16}$O \cite{jt740}, isotopologues of H$_3^+$
\cite{13FuSzMa.marvel,13FuSzFa.marvel}, and a series of investigations on
isotopologues of water \cite{jt454,jt482,jt539,jt576,jt562,jt750,jtwaterupdate}.
Note that the water studies provided
the original motivation for developing the  \Marvel\ procedure.
%The convention for naming data sources, tags which start with two digits 
%for the year of publication followed by the first two letters of author surnames,
%adopted for the work on water is also used here.

\subsection{Quantum numbers}
\Marvel\ requires that all transitions are assigned with a unique set of
descriptors, usually
quantum numbers, which are self-consistent across the entire dataset.
Many studies considered here used Hund's case $b$ quantum numbers for NH
and these descriptors were adopted for this work.
Each transition is therefore
labeled by an electronic state label (\X, \A, \Sa, or \Sc), a
vibrational quantum number ($v$), the total angular momentum quantum number ($J$),
the rotational angular momentum quantum number ($N$),
and parity ($e$/$f$) \cite{75BrHoHu.diatom}.  
For singlet states $N$ is set equal to $J$.

Figure 1 gives a schematic representation of the rovibronic band
systems considered in this work; it also shows the spin ($\Sigma$) and
orbital ($\Lambda$) angular momentum projections considered for each state.  
In the Hund's case $b$ representation employed here these
quantum numbers are not actually used.

In constructing a unified set of quantum numbers the following changes were made.
The label $F$ adopted for components of the \X\  state
by Bernath and co-workers \cite{82BeAm,86BrRaBe,99RaBeHi,96RaBe,10RaBe}
were mapped to the Hund's case $b$ quantum number $N$ as follows:

\vspace{-.5\baselineskip}
\begin{equation}
N=
\begin{cases}
J-1, & \text{if } F=1 \text{ and } p=e \\
J, 	 & \text{if } F=2 \text{ and } p=f \\
J+1, & \text{if } F=3 \text{ and } p=e \\
\end{cases}
\end{equation}

\noindent  The  parity of a rovibronic level incident to a 
rovibronic transition must satisfy one of the following relations:

\vspace{-1.0\baselineskip}
\begin{flalign}
\begin{aligned}
\Delta J = 1: & \ e \leftrightarrow e \text{ or } f \leftrightarrow f, \\
\Delta J = 0: & \ e \leftrightarrow f.
\end{aligned}
\end{flalign}

\noindent The interested reader should consult Figure~1 of Dixon \cite{59Dixon}
for an explanation how the various notations represent the transitions
within the \A--\X\  band system.

None of the studies considered here were able to resolve the
$\Lambda$-doubling of the \Sa\ state where each $(v,J=N)$ state should
appear with either $e$ or $f$ parity.
We decided to treat these
$(v,J=N)$ combinations as a single state and to adopt a parity label
of $d$ (for degenerate) in each case.
This simplification means that
odd-numbered cycles become possible in the experimental SN of $^{14}$NH.
In fact, while the great majority of the cycles within the
minimum cycle basis \cite{17ToFuCs} of the experimental SN of
$^{14}$NH have cycles of length four, the remaining cycles are largely
split equally between cycles with length three and five.
An example is the cycle formed by the transitions
27\,550.1873 \cite{86RaBe} and 30\,826.5245 \cm\ \cite{86RaBe} connecting
the \Sa\ and \Sc\ states with the same upper rovibronic level,
whereby the two lower rovibronic
states are connected by the measured transition 3276.3820 \cm\   \cite{85HaAdKaCu}.
Validation of these odd-numbered cycles supports
the assertion that $\Lambda$-doubling in the \Sa\ state is indeed small.

\section{Compilation of experimental sources}
$^{14}$NH has a pronounced hyperfine structure due to both $^{14}$N ($I=1)$ and H
$(I=\frac{1}{2}$) nuclear spins and most studies
of its pure rotational spectrum resolve this structure. 
However, as none of the studies at higher wavenumbers resolve the hyperfine structure,
it was decided not to consider hyperfine effects and treat the
rotational spectra hyperfine unresolved.
The hyperfine structure was therefore removed from sources 04LeBrWiSi \cite{04LeBrWiSi},
82VaMeDy \cite{82VaMeDy}, and 04FlBrMaOd \cite{04FlBrMaOd}.
Studies that only consider hyperfine transitions within a given $(v,J,N)$ state
\cite{84UbMeDy,87VaDaBrEv} were excluded  from our analysis at this stage.
Averaging the hyperfine components, of
course, leads to some loss of accuracy in the pure rotational levels
for which it should be possible to obtain hyperfine-resolved energies.
A number of older sources \cite{75RaLi,76WaRa,75MaGiVe,69KrNa}
were found to provide less accurate data than available from more
recent measurements. These sources were not considered further.

Our preliminary analysis also showed that including the transition
data of 82RaSa \cite{82RaSa} caused conflicts within the SN.
In 82RaSa \cite{82RaSa} there are 70 measured lines in the \Sc--\Sa\ band
system in the range 27\,105--27\,697 \cm\ with a claimed uncertainty of
about 0.005 \cm.
Nearly all these lines were remeasured as part of a
much more extended study by 86RaBe \cite{86RaBe}.
Removing six lines of 82RaSa \cite{82RaSa} from the singlet subnetwork allowed the 
SN to form correctly once the average uncertainty had been increased to 0.2 \cm.
We note that 82RaSa used a grating spectrograph and these measurements 
are essentially superseded by the  more accurate Fourier transform spectroscopy (FTS)  
measurements of 86RaBe; these two datasets agree once the uncertainty for 82RaSa is increased.
Remarkably, it was found necessary to remove only one further
line, due to 86BrRaBe \cite{86BrRaBe}, during MARVEL's validation process.

Infrared spectra of the \X\  state of $^{14}$NH are also available from two
distinct space-borne observations of solar spectra.
The ATMOS \cite{ATMOS} experiment, part of the Atmospheric Laboratory for
Applications and Science (ATLAS) Space Shuttle program,
provides solar spectra at a resolution of about 0.01 \cm.
The Atmospheric Chemistry Experiment (ACE) \cite{ACE2} is a Canadian
satellite mission which performs infrared solar occultation
observations of the Earth's atmosphere producing detailed solar
spectra in the infrared as a byproduct \cite{10HaWaMc}.
While the resolution of these spectra is lower than can be obtained in the
laboratory, typically 0.01 \cm\ or lower, the high temperature of the
Sun means that they probe rotational states not easily accessible in
laboratory experiments.
91GeSaGrFa \cite{91GeSaGrFa} used ATMOS to study pure rotational transitions 
of NH in the ground and first vibrationally excited states
spanning levels with $N$ from 20 to 42.
90GrLaSaVa \cite{90GrLaSaVa} provides ATMOS $R$-branch spectra of the
vibrational fundamental and the first hot band.
However, these spectra do
not resolve the fine structure splittings and there is only a single
transition, the fundamental $R(N''=27)$ line, at 3389.147(10) \cm, which
was not available from much higher resolution laboratory studies.
As a result, we decided not to include transitions from 90GrLaSaVa \cite{90GrLaSaVa}
in our final compilation.
Conversely, 10RaBe \cite{10RaBe} analyses ACE spectra in the region of the 
vibrational fundamental where ACE gives much higher signal-to-noise ratio than ATMOS.
10RaBe also present laboratory spectra, so we divided their results
into two segments and label them 10RaBe and 10RaBe\_S2 to distinguish these independent
subsets of data, with the latter source segment containing the less accurately
measured transitions.

\setlength{\tabcolsep}{3pt}
\renewcommand{\arraystretch}{1.25}
\setlength{\tabcolsep}{4pt} %increase the distance among columns
\captionof{table}{Data source segments and some of their characteristics 
for the $^{14}$NH molecule$^{a}$} %caption of the segment table
\addtocounter{table}{-1} %decrease counter of the tables by one as "\captionof" increased it by one
\vspace{-.5\baselineskip} %reduce space between caption and table
\scriptsize
\begin{longtable}{lcr@{--}p{1.0cm}>{\centering\arraybackslash}p{2.0cm}lll}
\hline \hline
\endfirsthead
\mc{6}{c}{\tablename\ \thetable\ -- \emph{Continued from previous page}}\\ \hline
     Segment tag                  &Band system  &  \mc{2}{c}{Range}         &        $A/V$      &  ESU       &  ASU       &  MSU       \\ \hline
\endhead
\hline \mc{6}{r}{\emph{ Continued on next page }} \\
\endfoot
\hline \hline
\endlastfoot
     Segment tag                  & Band system   &  \mc{2}{c}{Range}         &        $A/V$      &  ESU       &  ASU       &  MSU       \\ \hline
     97KlTaWi \cite{97KlTaWi}     &  \X\ -- \X    &  31.570     &  33.356     &         3/3       &  3.00e-06  &  3.00e-06  &  3.00e-06  \\
     04FlBrMaOd \cite{04FlBrMaOd} &  \X\ -- \X    &  65.213     &  162.67     &        12/12      &  5.00e-06  &  5.00e-06  &  5.00e-06  \\
     07RoBrFlZi \cite{07RoBrFlZi} &  \X\ -- \X    &  28.977     &  156.21     &        34/34      &  1.00e-05  &  1.00e-05  &  1.00e-05  \\
     82VaMeDy \cite{82VaMeDy}     &  \X\ -- \X    &  32.505     &  33.356     &         2/2       &  2.00e-05  &  7.54e-05  &  1.19e-04  \\
     86LeEvBr \cite{86LeEvBr}     &   \X\ -- \X   &  98.495     &  98.495     &         1/1       &  2.00e-05  &  2.00e-05  &  2.00e-05  \\
     04LeBrWiSi \cite{04LeBrWiSi} &   \X\ -- \X   &  32.505     &  66.169     &         4/4       &  1.00e-04  &  1.84e-04  &  3.99e-04  \\
     86BoBrChGu \cite{86BoBrChGu} &   \X\ -- \X   &  2310.1     &  3456.8     &       310/310     &  5.00e-04  &  7.82e-04  &  3.83e-03  \\
     82BeAm \cite{82BeAm}         &   \X\ -- \X   &  2950.0     &  3293.1     &        29/29      &  1.00e-03  &  1.64e-03  &  5.28e-03  \\
     85HaAdKaCu \cite{85HaAdKaCu} &   \Sa\ -- \Sa  &  3177.9     &  3537.2     &        35/35      &  1.00e-03  &  3.86e-03  &  3.88e-02  \\
     86BrRaBe \cite{86BrRaBe}     &  \A\ -- \X    &  26289      &  33080      &      1239/1238    &  1.00e-03  &  3.48e-03  &  4.11e-02  \\
     91GeSaGrFa \cite{91GeSaGrFa} &  \ X\ -- \X   &  622.45     &  941.28     &       103/103     &  1.00e-03  &  4.16e-03  &  2.80e-02  \\
     99RaBeHi \cite{99RaBeHi}     &  \X\ -- \X    &  2151.9     &  3459.0     &       500/500     &  1.00e-03  &  2.34e-03  &  2.63e-02  \\
     10RaBe \cite{10RaBe}         &  \X\ -- \X    &  2151.8     &  2530.2     &        36/36      &  1.00e-03  &  1.00e-03  &  1.00e-03  \\
     10RaBe\_S2 \cite{10RaBe}     &   \X\ -- \X   &  2662.1     &  3459.0     &       266/266     &  5.00e-02  &  5.00e-02  &  5.00e-02  \\
     86UbMeTeDy \cite{86UbMeTeDy} &  \Sc\ -- \Sa  &  30059      &  30742      &        38/38      &  2.00e-03  &  5.57e-03  &  1.17e-02  \\
     03VaSaMoJo \cite{03VaSaMoJo} & \A\ -- \Sa    &  17002      &  17193      &        52/52      &  2.00e-03  &  1.10e-02  &  2.74e-02  \\
     86RaBe \cite{86RaBe}         &   \Sc\ -- \Sa  &  26653      &  32938      &       250/250     &  3.00e-03  &  4.57e-03  &  3.85e-02  \\
     82RaSa \cite{82RaSa}         &  \Sc\ -- \Sa  &  27106      &  27697      &        70/64      &  5.00e-03  &  2.64e-02  &  6.98e-02  \\
     90HaMi \cite{90HaMi}         &  \Sc\ -- \Sa  &  21608      &  24525      &        18/18      &  2.00e-02  &  2.00e-02  &  2.00e-02  
\end{longtable}
\normalsize
\normalsize
\vspace{-.5\baselineskip} %reduce space between table and footnote
\makeatletter\def\@currentlabel{\thetable}\label{table:segTable} %cross reference for the segment table
{\setstretch{1.0}
\small
\noindent $^{a}$ Tags denote experimental data-source segments employed during this study
(the identifier associated with the first segment of a data source, `\_S1',
is not written out explicitly).
The column `Range' indicates the range (in \cm) corresponding to validated
wavenumber entries within the experimental linelist.
`$A/V$' is an ordered pair, where $A$ and $V$ are the number of assigned and validated
transitions related to a given source segment, respectively, obtained at the end of
the MARVEL analysis.
ESU, ASU, and MSU designate the estimated, the average,
and the maximum segment uncertainties in \cm, respectively.
Rows of this table are arranged in the order of
the ESUs with the restriction that the segments of the same data source should be listed
one after the other.\\
\normalsize
\par}

\setstretch{1.5}

There are a number of studies of the intercombination band system
which link the singlet and triplet subnetworks \cite{84RoSt,99RiGe,03VaSaMoJo}.
99RiGe \cite{99RiGe} used stimulated emission pumping of the strongly forbidden
\Sa\ -- \X\ band system to study the singlet-triplet splitting.
However, the data of 99RiGe \cite{99RiGe} only have an accuracy of 0.1 \cm\   or
worse and the source does not actually provide the primary transition data,
so this source was not considered further.
03VaSaMoJo \cite{03VaSaMoJo} used optical
pumping to study \Sa--\A\  transitions with an accuracy of 0.03 \cm\  or better. 
03VaSaMoJo \cite{03VaSaMoJo} also used the F labelling convention; therefore,
their assignments were transformed as described above.
Lower  resolution experimental studies of the \Sa--\X\  emission spectra by 
84RoSt \cite{84RoSt} do not provide any transition wavenumbers and were not
considered further.

\begin{figure}
\centering
\includegraphics[width=0.65\textwidth]{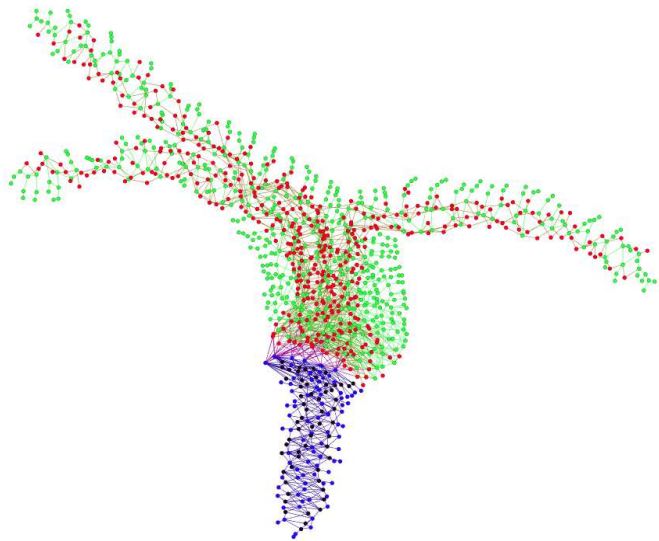}
\caption{\label{fig:SN} The 
experimental spectroscopic network of the $^{14}$NH molecule showing the different
electronic states considered during this work with different color,
red: \X, green: \A, blue: \Sa, and black: \Sa.
The relatively large number of bridges characterizing the spectroscopic network of
$^{14}$NH is evident from this figure.}
\end{figure}

\renewcommand{\arraystretch}{1.5}
\section{Results and Discussion}

Table~\ref{table:segTable} presents a summary of the transitions data
used in the final \Marvel\ run of this study.
The latest version of the MARVEL code (\texttt{intMARVEL} \cite{jt750}),
with some minor modifications, was 
used to obtain the energy levels and the refined transitions.
A total of \NoTR\  assigned transitions from \NoSO\ distinct data sources
were included in our final \Marvel\ analysis.
Of these transitions and energy
levels, \NoPrCoTR\ and \NoPrCoEL, respectively, are contained within
the principal component of the experimental SN of $^{14}$NH linking
levels in all four electronic states of interest.  
\NoFlCoTR\    transitions and \NoFlCoEL\ energy levels could not be linked to the
principal component, they are part of floating components, but they
are retained in the dataset as they may be linked to the principal
component when new experimental data become available.

After the reassignments and other manipulations considered in the last section,
only \NoExclTR\ transitions had to be removed from the dataset considered
by \Marvel\ as not consistent with the other transitions.
These transitions are retained in the final transition list but are given
as negative wavenumber entries.

The principal component of the experimental SN of $^{14}$NH assembled during this study
links together \NoPrCoEL\  energy levels (see Figure~\ref{fig:SN}).
Of these energy levels, \NoPpEL, \NoPmEL, \NoSEL, \NoUEL\  have resistances 
P$^+$, P$^-$, S, and U, respectively (definitions of the labels can be
found in the ReadMe.txt file of the Supplementary Material).
Energy levels with resistance P$^+$ are indeed dependable,
while the others may change when even more accurate lines will be included 
in the database of experimental transitions.
Although it is likely that even the non-P$^+$ energy
levels are correct, especially those of P$^-$, they cannot be regarded as fully verified by the present analysis.

\begin{figure}
\centering
\includegraphics[width=0.75\textwidth]{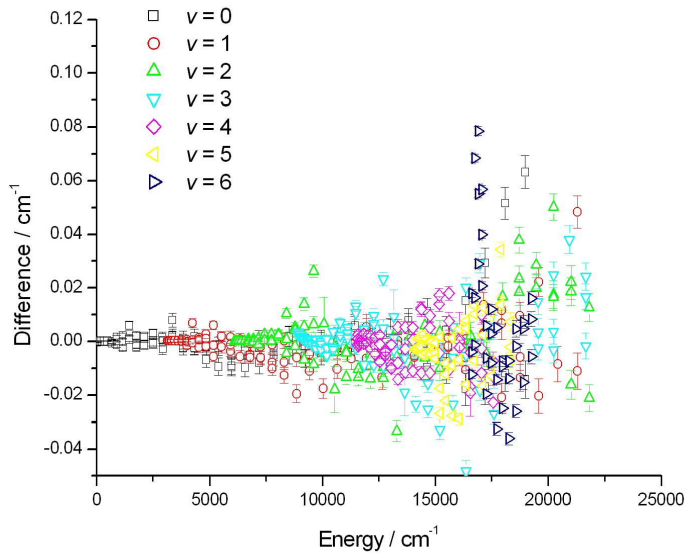}
\caption{\label{fig:diffX}Differences between the empirical MARVEL energy levels 
of this study and the earlier literature results of 15BrBeWe \cite{15BrBeWe.NH}
related to the \X\ electronic state.}
\end{figure}

Due to the lack of truly high-accuracy first-principles energies for $^{14}$NH,
we used spectroscopic constants available in the literature to confirm the empirical
(MARVEL) rovibrational energy levels derived during this study.
For the \X\ state,
we used the PGOPHER \cite{PGOPHER} file of 15BrBeWe \cite{15BrBeWe.NH} to compute
the rovibrational energy levels.
Figure~\ref{fig:diffX} shows the differences between the MARVEL levels
and this earlier work.
As seen in Figure~\ref{fig:diffX}, there are only few energy levels where the difference
is larger than 0.05 \cm\, so the agreement is considered to be very good.

\begin{table}[t!]
\caption{Focal-point-analysis table of the $T_{\rm e}$ excitation energy between
the \X\  and  \Sa\  states of $^{14}$NH.}
\label{table:FPA}
\resizebox{\columnwidth}{!}{%
\begin{tabular}{lrrrrrrrrrrrrrrrrr}
\hline
\hline
frozen-core &$\Delta E_{\rm e}$(UHF)&$\delta$[MP2]&$\delta$[CCSD]&$\delta$[CCSD(T)]&$\delta$[CCSDT]&$\delta$[CCSDTQ]&$\delta$[CCSDTQP]&$\delta$[CCSDTQPH]&$\Delta E_{\rm e}$[FCI] \\ \hline
%aug-cc-pVDZ & &		\\
aug-cc-pwCVDZ   & 23\,300.4&--3981.3&--2229.6&--1079.7&--1787.2&--148.6  &--4.3  &--0.0&14\,069.9	\\
aug-cc-pwCVTZ   & 23\,222.7&--4838.1&--2043.0&--1078.9&--2002.3&--229.9  &[--4.3]&[--0.0]&[11\,947.3] \\
aug-cc-pwCVQZ   & 23\,236.1&--5168.2&--1914.7&--1074.6&--2034.0&--247.7  &[--4.3]&[--0.0]&[11\,735.8] \\
aug-cc-pwCV5Z   & 23\,240.9&--5316.0&--1836.4&--1074.9&--2039.4&--~~~    &[--4.3]&[--0.0]&[11\,665.0] \\
%aug-cc-pV6Z    & 15672.5    & 554.9& 235.4& 16462.8	\\
CBS           &$\mathbf{23\,241.7(20)}$&$\mathbf{-5471.1(500)}$&$\mathbf{-1754.2(250)}$&$\mathbf{-1075.3(10)}$&$\mathbf{-2045.1(40)}$&$\mathbf{-266.5(250)]}$&$\mathbf{[-4.3(20)]}$&$\mathbf{[-0.0(10)]}$&$\mathbf{12\,625.2(750)}$ \\ \hline
all-electron&$\Delta E_{\rm e}$(UHF)&$\delta$[MP2]&$\delta$[CCSD]&$\delta$[CCSD(T)]&$\delta$[CCSDT]&$\delta$[CCSDTQ]&$\delta$[CCSDTQP]& &$\Delta E_{\rm e}$ \\ \hline
%aug-cc-pVDZ & &		\\
aug-cc-pwCVDZ   & 23\,300.4&--3995.9&--2133.8&--1093.6&--1846.8&--159.8&--4.9  & &14\,065.6 \\
aug-cc-pwCVTZ   & 23\,222.7&--4823.6&--1924.2&--1089.4&--2100.5&--254.3&[--4.9]&  \\
aug-cc-pwCVQZ   & 23\,236.1&--5149.6&--1794.2&--1083.1&--2136.0&--276.0&[--4.9]&  \\
aug-cc-pwCV5Z   & 23\,240.9&--5296.1&--1716.1&--1082.6&--~~~   &--~~~  &[--4.9]& \\
%aug-cc-pV6Z    & 15672.5    & 554.9& 235.4& 16462.8	\\
CBS           &$\mathbf{23\,241.7(20)}$&$\mathbf{-5449.9(500)}$&$\mathbf{-1634.1(250)}$&$\mathbf{-1082.0(20)}$&$\mathbf{-2168.0(40)}$&$\mathbf{-298.7(250)}$&$\mathbf{[-4.9(20)]}$& &$\mathbf{12\,604.1(750)}$  \\
\hline
\hline
\end{tabular}}
\noindent $^a$ The symbol $\delta$ denotes the increment in the relative energy
($\Delta E_{\rm e}$) with respect to the preceding level of theory in the hierarchy
HF $\rightarrow$ MP2 $\rightarrow$ CCSD $\rightarrow$ CCSD(T) $\rightarrow$ CCSDT $\rightarrow$ CCSDTQ $\rightarrow$ CCSDTQP $\rightarrow$ CCSDTQPH
($\equiv$ FCI in the case of the frozen-core approximation).
CBS = complete basis set.
The basis set extrapolations are 
%described in the text, they are 
based on the cardinal number $X$ of the aug-cc-pCV$X$Z Gaussian basis-set family,
for electron correlation increments they follow the formula $X^{-3}$ and employ
the largest two $X$ values available.
Uncertainties are given in parentheses.
All energy values are given in \cm.
\end{table}

For the energy levels of the \Sa\  state, we used the molecular constants of
86RaBe \cite{86RaBe}, in the cases of $v=0$ and 1, and 90HaMi \cite{90HaMi},
for $v=2$ and 3.
Prior to 1974 \cite{74GiMaVe}, no intercombination transitions could be measured;
therefore, the energy separation of the singlet and triplet states of NH had
to be estimated by indirect experimental data or by \textit{ab initio} calculations.
This is the reason why in 1963 McBride \cite{63McHeEhGo} set the value of
$T_{\rm e}$(\Sa) to 14\,922 \cm\ but in 1979, in their famous book,
Huber and Herzberg \cite{79HuHe} published 12\,566 \cm\ for this value,
and 10 years later Gurvich \cite{Gurvich2} used a value of
12\,577.1 \cm\  for $T_{\rm e}$ when calculating the thermochemical functions
of the $^{14}$NH molecule.
The large shift in the $T_0$ value of the \Sa\  state over time can be explained by
the pronounced multireference character of this state, due partly to the fact that
it dissociates to an excited state of the N atom.
The considerable difficulties the gold standard CCSD(T) level has in providing
a correct estimate of $T_{\rm e}$(\Sa) can be seen from the frozen-core and 
all-electron focal-point-analysis-type \cite{93AlEaCs,98CsAlSc} entries of Table~\ref{table:FPA}.
The CCSD to CCSDT increment is huge and CCSD(T) can recover only less than half
of this increment, a rather unusual situation.
Though the CCSDTQ increment is not particularly large, it is unusually basis-set
dependent, showing the necessity of a multireference treatment for obtaining the
excitation energy of the \Sa\  state.
The increments above CCSDTQ are very small, as usual and expected.
Overall, the excitation energy is only about half of its unrestricted Hartree--Fock (UHF)
estimate.
One can estimate $T_{\rm e}$(\Sa) starting from the all-electron CBS CCSDTQ value
of Table~\ref{table:FPA}, 12\,604(75) \cm, and adding to this the 
higher-order coupled-cluster estimates, $-5(2)$ \cm, as well as the relativistic
correction (computed at the all-electron aug-cc-pwCVQZ CCSD(T) level), $-2$ \cm.
Adding the zero-point vibrational energy (ZPVE) correction, +27(5) \cm,
obtained again at the aug-cc-pwCVQZ CCSD(T) level,
yields our final $T_{\rm e}(T_{\rm 0}$) estimates of 12\,597(12\,624) \cm,
with a conservative $2\sigma$ uncertainty estimate of $\pm 80$ \cm.
These excitation energies support the best previous experimental estimates
mentioned above.

\begin{figure}
\centering
\includegraphics[width=0.75\textwidth]{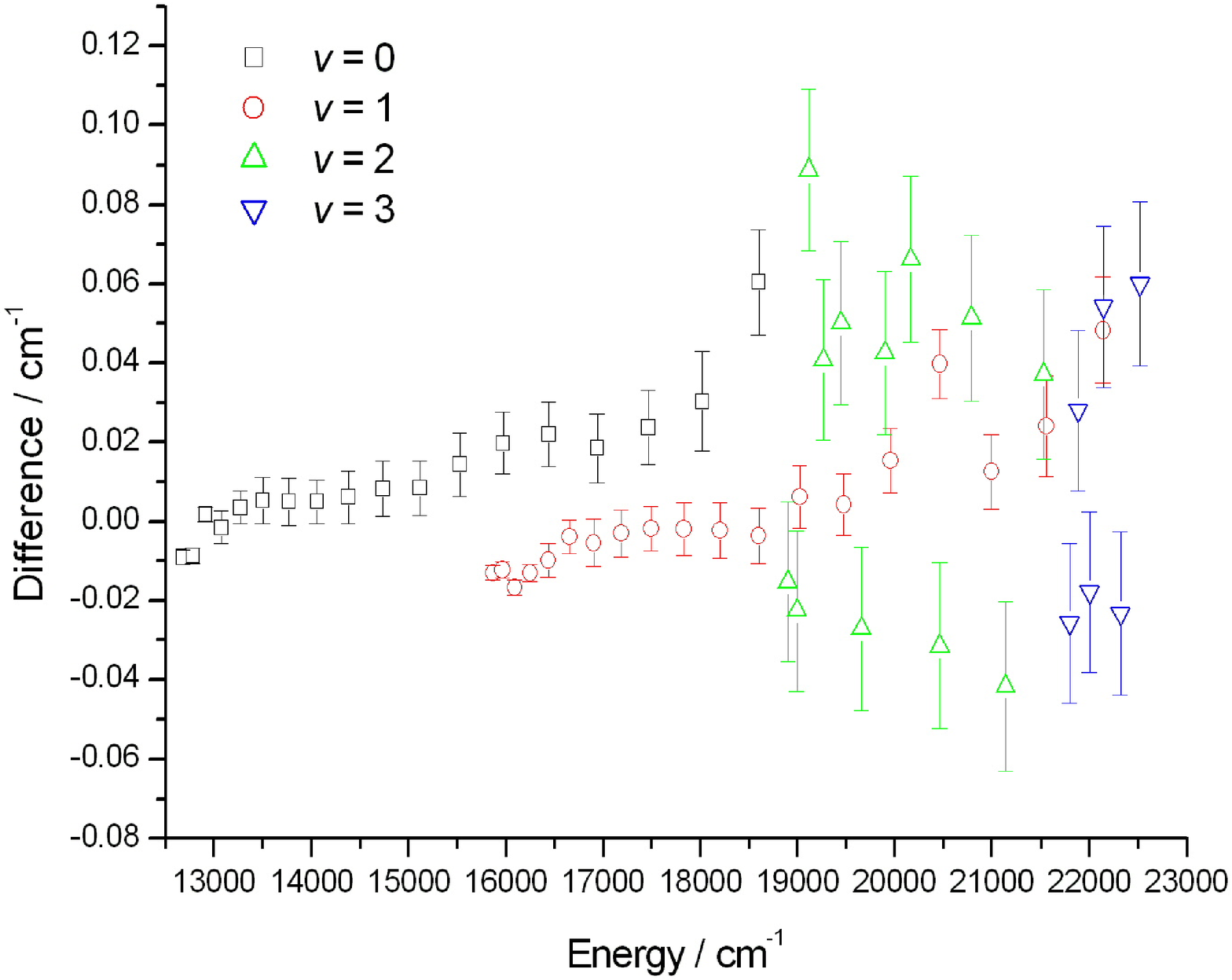}
\caption{\label{fig:diffa}Differences between the empirical MARVEL energy levels 
of this study and the earlier literature results of 86RaBe \cite{86RaBe} 
and 90HaMi \cite{90HaMi} related to the \Sa\ electronic state.}
\end{figure}

\begin{figure}
\centering
\includegraphics[width=0.75\textwidth]{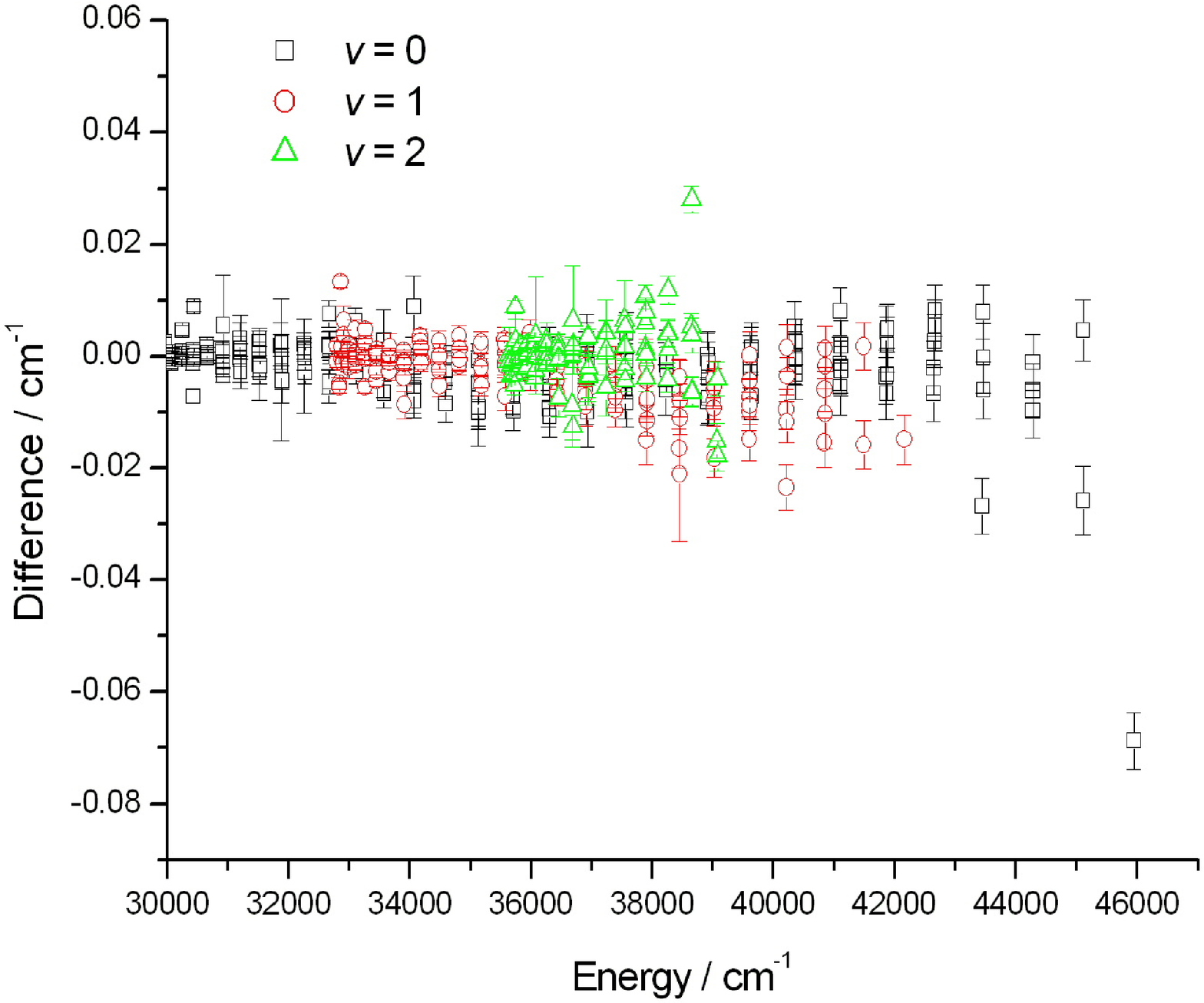}
\caption{\label{fig:diffA3}Differences between the empirical MARVEL energy levels 
of this study and the earlier literature results of 10RaBe \cite{10RaBe} 
related to the \A\ electronic state.}
\end{figure}

\begin{figure}
\centering
\includegraphics[width=0.75\textwidth]{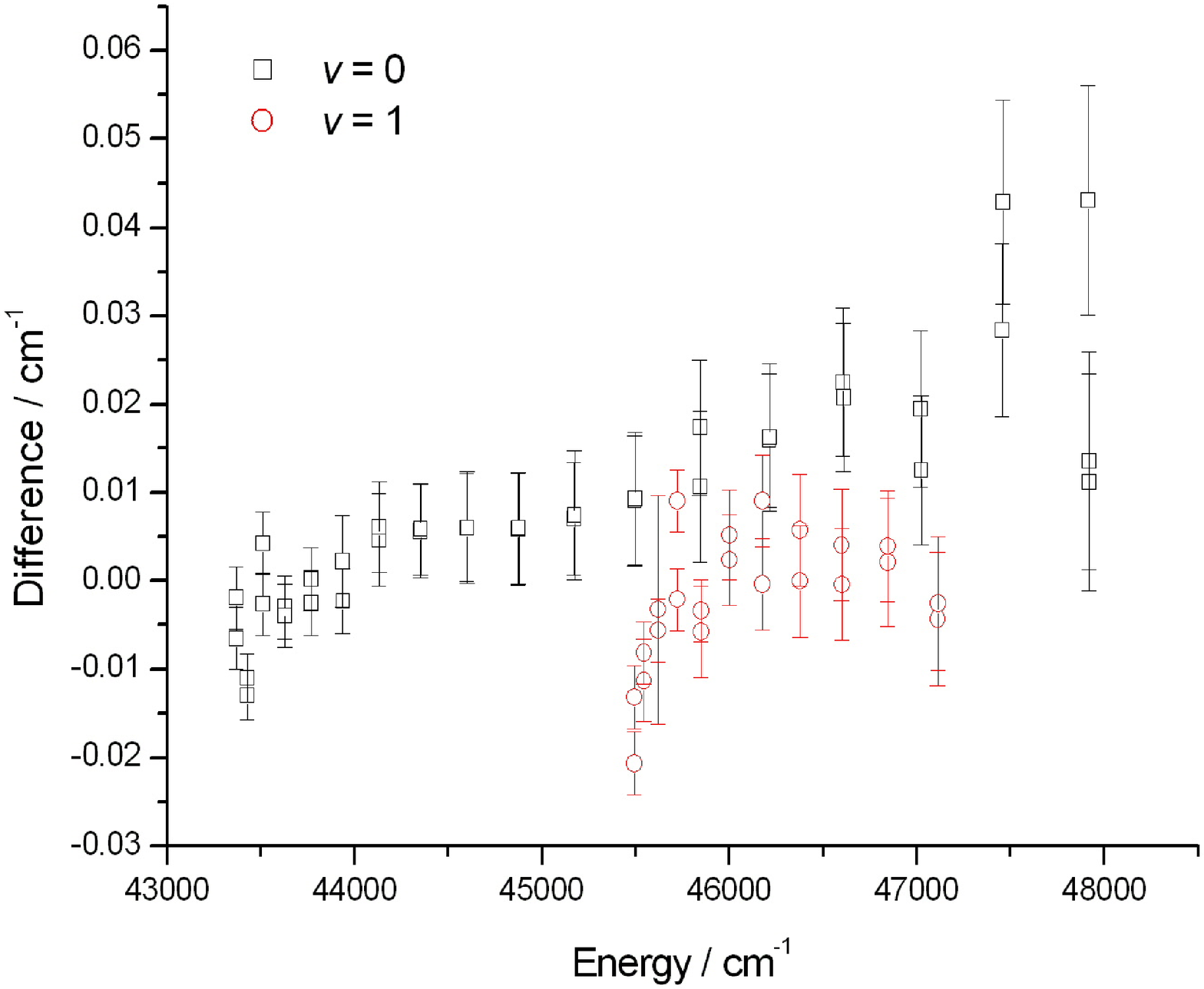}
\caption{\label{fig:diffc1}Differences between the empirical MARVEL energy levels 
of this study and the earlier literature results of 86RaBe \cite{86RaBe} related to 
the \Sc\ electronic state. }
\end{figure}

The first high-accuracy measurement of the singlet-triplet splitting,
between the \Sa\ ($v=0$, $J=N=2$) and \X\ ($v=0$, $J=1$, $N=0$) states,
was obtained by 86RaBe \cite{86RaBe}, the value is 12\,688.39(10) \cm.
This energy difference was given as 12\,687.8(1) \cm\ in 99RiGe \cite{99RiGe}
and probably the best value is given by 03VaSaMoJo \cite{03VaSaMoJo}, 
it is 12 688.622(4) \cm.
The MARVEL value for this splitting is 12\,688.612(2) \cm.
Since the $T_{\rm e}$ values are not accurate enough to confirm the MARVEL levels
and there are no high-accuracy values for $T_v$ in the literature,
we fitted a simple Hamiltonian, taken from 86RaBe \cite{86RaBe}, for the degenerate
energy levels of the \Sa\  state.
We obtain the following values for $T_e+T_v$:
\datum{12\,655.761}, \datum{15\,838.558}, \datum{18\,875.218}, and
\datum{21\,768.236} \cm\   for $v=0$, 1, 2, and 3, respectively.
Using these $T_e+T_v$ values and the rotational constants of 86RaBe \cite{86RaBe} and 
90HaMi \cite{90HaMi}, we could check the accuracy of the MARVEL energy levels.
Figure~\ref{fig:diffa} shows the differences between the MARVEL levels and the 
effective Hamiltonian results.
It can be seen that for the $v=0$ and 1 vibrational states the agreement is nearly
perfect, but in the cases of higher vibrational states the differences usually are 
larger than 0.03 \cm.

For checking the MARVEL energy levels of the \A\ state, we used the term values 
of 10RaBe \cite{10RaBe}.
The result of the comparison of the present empirical and the effective Hamiltonian
energies can be seen in Figure~\ref{fig:diffA3}.
Only two MARVEL energy levels could not be reproduced within 0.03 \cm.

%\newpage
\begin{table}[t!]
\caption{Vibrational energy levels of $^{14}$NH.$^{a}$} %caption of the segment table
%\captionof{table}{Vibrational energy levels of $^{14}$NH.$^{a}$} %caption of the segment table
%\addtocounter{table}{-1} %decrease counter of the tables by one as "\captionof" increased it by one
%\vspace{-.0\baselineskip} %reduce space between caption and table
%\begin{table}[t!]
%\begin{table}
\label{table:vibtab}
\begin{tabular}{lllllrr}
\hline \hline
ES  & $v$ & $J$ & $N$ & $p$ & $E$(MARVEL)/\cm            & $E$(86BoBrChGu \cite{86BoBrChGu})/\cm \\
\hline
\X\ &  0  &  1  &  0  & $e$ & 0                          & 0                   									 \\
\X\ &  1  &  1  &  0  & $e$ & \datum{3125.57241(50)}     & 3125.57291(16)      									 \\
\X\ &  2  &  1  &  0  & $e$ & \datum{6094.87429(71)}     & 6094.87502(20)      									 \\
\X\ &  3  &  1  &  0  & $e$ & \datum{8907.59729(87)}     & 8907.59833(25)      									 \\
\X\ &  4  &  1  &  0  & $e$ & \datum{11562.3129(13)}     & 11562.31434(42)										   \\
\hline\hline
\end{tabular}
\end{table}
\vspace{-.5\baselineskip} %reduce space between table and footnote
{\setstretch{1.0}
\small
\noindent $^{a}$ ES = electronic state.
The uncertainties of the reference data are calculated as the
square root of the squared sum of the uncertainties found in Table 1 of Ref. \cite{86BoBrChGu}.
\\
\par
\normalsize}

\setstretch{1.5}

We used the spectroscopic constants of 86RaBe \cite{86RaBe} and our $T_v$ values
for the \Sc\  state to compute the energy levels of the \Sc\  state.
Figure~\ref{fig:diffc1} shows the result of this comparison:
the agreement is almost perfect.

The list of MARVEL energy levels and the corresponding experimental transitions are 
placed in supplementary data. 

Table~\ref{table:vibtab} compares the vibrational band origins
obtained during the MARVEL analysis with available experimental
results.  It is striking that direct vibrational band origin information is
available only for the ground electronic state of $^{14}$NH and only
for $v=1$-4 even though some experimental rovibrational data are available for $v$
up to 6 in the \X\ state, up to 3 in the \Sa\ state, up to 2 in the \A\ state,
and for $v=0$ and 1 in the \Sc\ state.  
It is also worth noting that the uncertainties obtained during the MARVEL analysis
are slightly larger than the $1\sigma$ values reported in 86BoBrChGu \cite{86BoBrChGu}.
Nevertheless, the agreement between MARVEL and 86BoBrChGu is excellent.

\section{Conclusions}
\label{sec.diss}

A \Marvel\ analysis of the observed and assigned rovibronic transitions
has been performed for the parent imidogen radical, $^{14}$NH.
After making a number of minor adjustments,
a set of \NoPrCoEL\  empirical rovibronic energy levels related
to the principal component of the SN, containing \NoPrCoTR\  validated transitions,
are obtained for this radical spanning four low-lying electronic states.
If desired, these energies provide suitable
input for an effective Hamiltonian fit.

There are some high-resolution spectroscopic data available on
higher-lying electronic states of NH including the \Sd\  state \cite{78GrLe,91AsClHo}
and various Rydberg states \cite{92ClAsWe,92ClAsWeJo,92ClAsWeDe}.
However, at this stage there are insufficient data available for any of these
states to make their inclusion in a \Marvel\ analysis productive.
The $^{14}$NH data produced here will be included in the newly formed \Marvel\
project database \cite{jtMARVEL} and can be actively updated should
new high-resolution assigned spectra become available.

Finally, we note that a significant part of this work was performed by
pupils from Preston Manor School in north-west London, as part of the
ORBYTS (Original Research By Young Twinkle Students) project.
Several other \Marvel\  studies have been completed as part of the ORBYTS project,
including those on $^{48}$Ti$^{16}$O \cite{jt672}, acetylene \cite{jt705},
H$_2$S \cite{jt718}, $^{90}$Zr$^{16}$O \cite{jt740}, and methane \cite{jtCH4Marvel}.
Sousa-Silva {\it et al.} \cite{jt709} discusses our experiences of working with
high-school students on  high-level research projects.
High-school students in Hungary were also involved in a \Marvel\ 
analysis of the high-resolution spectra of $^{16}$O$_2$ \cite{19FuHoKoSo}.

%\begin{acknowledgement}
\section*{Acknowledgement}
We thank Dr. Ehsan Pedram of Preston Manor School for his considerable
help during the course of this project.
This work was supported  by the Engineering and Physical Sciences Research
Council (EPSRC) for a studentship for DD-L under grant EP/M507970/1.  
The work performed in Budapest received support from NKFIH (grant no. K119658)
and from the ELTE Excellence Program (1783-3/2018/FEKUTSTRAT) supported by
the Hungarian Ministry of Human Capacities (EMMI).
The collaboration between the Budapest and London groups received support from
the European Cooperation in Science and Technology (COST) action
CM1405, MOLIM: Molecules in Motion.
Some support was provided by the NASA Laboratory Astrophysics program. 

\bibliographystyle{elsarticle-num}

%\bibliography{journals_phys,jtj,NH-MARVEL,MARVEL,NH,methods,diatomic,SO2,atmos,NHextra}

\begin{thebibliography}{100}
\expandafter\ifx\csname url\endcsname\relax
  \def\url#1{\texttt{#1}}\fi
\expandafter\ifx\csname urlprefix\endcsname\relax\def\urlprefix{URL }\fi
\expandafter\ifx\csname href\endcsname\relax
  \def\href#1#2{#2} \def\path#1{#1}\fi

\bibitem{75MaGiVe}
J.~Masanet, A.~Gilles, C.~Vermeil, Light emission of the photofragments
  produced by photolysis of ammonia and ammonia-$d_3$ at 147, 123.6 and 104.8
  nm: First observation of the {b\,$^1\Sigma^+$ -- X\,$^3\Sigma^-$} transition
  of {NH} and {ND}, J. Photochem. 3 (1974) 417--429.

\bibitem{85MaHiMo}
Y.~Maruyama, T.~Hikida, Y.~Mori, {Formation of NH (A $^3\Pi_i$) in the flash
  photolysis of HN$_3$ at 121.6 nm. Role of N$_2$ triplet states}, Chem. Phys.
  Lett. 116 (1985) 371--373.

\bibitem{18HaBrPh.NH}
S.~S. Harilal, B.~E. Brumfield, M.~C. Phillips, {An evaluation of equilibrium
  conditions and temperature-dependent speciation in a laser-produced air
  plasma}, Phys. Plasmas {25} ({2018}) 083303 .
%\newblock \href {https://doi.org/{10.1063/1.5041987}}  {\path{doi:{10.1063/1.5041987}}}.

\bibitem{17PfOuBe.NH}
R.~Pflieger, T.~Ouerhani, T.~Belmonte, S.~I. Nikitenko, {Use of NH (A~$^3\Pi$
  -- X~$^3\Sigma^-$) sonoluminescence for diagnostics of nonequilibrium plasma
  produced by multibubble cavitation}, Phys. Chem. Chem. Phys. {19} ({2017})
  26272--26279.
%\newblock \href {https://doi.org/{10.1039/c7cp04813k}} {\path{doi:{10.1039/c7cp04813k}}}.

\bibitem{18HaLiCh.NH}
A.~Hamdan, J.-L. Liu, M.~S. Cha, {Microwave plasma jet in water:
  Characterization and feasibility to wastewater treatment}, Plasma Chem.
  Plasma Proc. {38} ({2018}) 1003--1020 .
%\newblock \href {https://doi.org/{10.1007/s11090-018-9918-y}}  {\path{doi:{10.1007/s11090-018-9918-y}}}.

\bibitem{19ZhGaLi.NH}
D.~Zhang, Q.~Gao, B.~Li, J.~Liu, Z.~Li, {Ammonia measurements with femtosecond
  laser-induced plasma spectroscopy}, Appl. Optics {58} ({2019}) 1210--1214.
%\newblock \href {https://doi.org/{10.1364/AO.58.001210}}  {\path{doi:{10.1364/AO.58.001210}}}.

\bibitem{18PeChAk.NH}
R.~Perillo, R.~Chandra, G.~R.~A. Akkermans, W.~A.~J. Vijvers, W.~A. A.~D.
  Graef, I.~G.~J. Classen, J.~van Dijk, M.~R. de~Baar, {Studying the influence
  of nitrogen seeding in a detached-like hydrogen plasma by means of numerical
  simulations}, Plasma Phys. Controlled Fussion {60} ({2018}) 105004 .
%\newblock \href {https://doi.org/{10.1088/1361-6587/aad703}}  {\path{doi:{10.1088/1361-6587/aad703}}}.

\bibitem{18PaDiDr.NH}
E.~Pawelec, T.~Dittmar, A.~Drenik, A.~Meigs, J.~Contributors, {Molecular ND
  Band Spectroscopy in the Divertor Region of Nitrogen Seeded JET Discharges},
  J. Phys. Conf. Ser. {959} ({2018}) 012009 .
%\newblock \href {https://doi.org/{10.1088/1742-6596/959/1/012009}}  {\path{doi:{10.1088/1742-6596/959/1/012009}}}.

\bibitem{18BrNiNa.NH}
C.~Brackmann, E.~J.~K. Nilsson, J.~D. Naucler, M.~Alden, A.~A. Konnov,
  {Formation of NO and NH in NH$_3$-doped CH$_4$ + N$_2$ + O$_2$ flame:
  Experiments and modelling}, Combust. Flame {194} ({2018}) 278--284 .
%\newblock \href {https://doi.org/{10.1016/j.combustflame.2018.05.008}}  {\path{doi:{10.1016/j.combustflame.2018.05.008}}}.

\bibitem{19LaGaDe.NH}
N.~Lamoureux, L.~Gasnot, P.~Desgroux, {Quantitative NH measurements by using
  laser-based diagnostics in low-pressure flames}, Proc. Combustion Inst. {37}
  ({2019}) 1313--1320.
%\newblock \href {https://doi.org/10.1016/j.proci.2018.09.007}  {\path{doi:10.1016/j.proci.2018.09.007}}.

\bibitem{07CaTsLu.NH}
W.~C. Campbell, E.~Tsikata, H.-I. Lu, L.~D. van Buuren, J.~M. Doyle, {Magnetic
  trapping and Zeeman relaxation of NH (X $^3\Sigma^-$)}, Phys. Rev. Lett. 98
  (2007) 213001.
%\newblock \href {https://doi.org/10.1103/PhysRevLett.98.213001}  {\path{doi:10.1103/PhysRevLett.98.213001}}.

\bibitem{10TsCaHu.NH}
E.~Tsikata, W.~C. Campbell, M.~T. Hummon, H.-I. Lu, J.~M. Doyle, {Magnetic
  trapping of NH molecules with 20 s lifetimes}, New J. Phys {12} ({2010})
  065028.
% \newblock \href {https://doi.org/{10.1088/1367-2630/12/6/065028}} {\path{doi:{10.1088/1367-2630/12/6/065028}}}.

\bibitem{11WaLoZu.NH}
A.~O.~G. Wallis, E.~J.~J. Longdon, P.~S. Zuchowski, J.~M. Hutson, {The
  prospects of sympathetic cooling of NH molecules with Li atoms}, Eur. Phys.
  J. D {65} ({2011}) 151--160.
%\newblock \href {https://doi.org/10.1140/epjd/e2011-20025-4}  {\path{doi:10.1140/epjd/e2011-20025-4}}.

\bibitem{13JaVaGr.NH}
L.~M.~C. Janssen, A.~van~der Avoird, G.~C. Groenenboom, {Quantum reactive
  scattering of ultracold NH(X $^3\Sigma^-$) Radicals in a magnetic trap},
  Phys. Rev. Lett. {110} ({2013}) 063201.
%\newblock \href {https://doi.org/{10.1103/PhysRevLett.110.063201}}  {\path{doi:{10.1103/PhysRevLett.110.063201}}}.

\bibitem{02RiGe.NH}
J.~L. Rinnenthal, K.-H. Gericke, {State-to-state energy transfer of NH (X
  $^3\Sigma^-$, $v= 0, J, N$) in collisions with He and N$_2$}, J. Chem. Phys.
  116 (2002) 9776--9791.

\bibitem{06Kajita.NH}
M.~Kajita, {Collision between magnetically trapped NH molecules in the ($N=0,
  J=1$) state}, Phys. Rev. A 74~(3) (2006) 032710.

\bibitem{79SmRo}
J.~H. Smith, D.~W. Robinson, {Pure rotational lasing in four electronic states
  of NH: Impulsive to adiabatic collisional pumping}, J. Chem. Phys. 71 (1979)
  271--280.

\bibitem{93Eder}
J.~M. Eder, Contributions to spectral analysis, Denkschr. Wien Akad. 60 (1893)
  1--24.

\bibitem{18FoGr.NH}
A.~Fowler, C.~C.~L. Gregory, {The ultra-violet band of ammonia, and its
  occurrence in the solar spectrum}, Proc. Roy. Soc. A {94} ({1918}) 470--471.
%\newblock \href {https://doi.org/10.1098/rspa.1918.0032}  {\path{doi:10.1098/rspa.1918.0032}}.

\bibitem{73GrSa}
N.~Grevesse, A.~J. Sauval, A study of molecular lines in the solar photospheric
  spectrum, Astron. Astrophys. 27 (1973) 29.

\bibitem{41SwElBa.NH}
P.~Swings, T.~C. Elvey, H.~W. Babcoc, {The spectrum of Comet Cunningham,
  1940c}, Astrophys. J. {94} ({1941}) 320--343.
%\newblock \href {https://doi.org/10.1086/144336} {\path{doi:10.1086/144336}}.

\bibitem{72LaBe.NH}
D.~L. Lambert, R.~Beer, {Vibration-rotation bands of NH in the spectrum of
  alpha Orionis}, Astrophys. J. 177 (1972) 541.

\bibitem{91MeRo.NH}
D.~M. Meyer, K.~C. Roth, Discovery of interstellar {NH}, Astrophys. J. {376}
  ({1991}) L49--L52.
%\newblock \href {https://doi.org/10.1086/186100} {\path{doi:10.1086/186100}}.

\bibitem{97CrWi.NH}
I.~A. Crawford, D.~A. Williams, {Detection of interstellar NH towards zeta
  Ophiuchi by means of ultra-high-resolution spectroscopy}, Mon. Not. R.
  Astron. Soc. {291} ({1997}) L53--L56.
%\newblock \href {https://doi.org/10.1093/mnras/291.3.L53}  {\path{doi:10.1093/mnras/291.3.L53}}.

\bibitem{09WeGaBe.NH}
T.~Weselak, G.~A. Galazutdinov, Y.~Beletsky, J.~Krelowski, {Interstellar NH
  molecule in translucent sightlines}, Mon. Not. R. Astron. Soc. {400} ({2009})
  392--397.
%\newblock \href {https://doi.org/10.1111/j.1365-2966.2009.15466.x} {\path{doi:10.1111/j.1365-2966.2009.15466.x}}.

\bibitem{97AoTs.NH}
W.~Aoki, T.~Tsuji, {High resolution infrared spectroscopy of CN and NH lines:
  nitrogen abundance in oxygen-rich giants through K to late M}, Astron.
  Astrophys. {328} ({1997}) 175--186.

\bibitem{34DiBl}
G.~H. Dieke, R.~W. Blue, {A
  $^{1}\ensuremath{\Pi}\ensuremath{\rightarrow}^{1}\ensuremath{\Delta}$ Band of
  NH and the Corresponding ND Band}, Phys. Rev. 45 (1934) 395--400.
%\newblock \href {https://doi.org/10.1103/PhysRev.45.395} {\path{doi:10.1103/PhysRev.45.395}}.

\bibitem{14BrBeWe.NH}
J.~S.~A. Brooke, P.~F. Bernath, C.~M. Western, M.~C. {van Hemert}, G.~C.
  Groenenboom, {Line strengths of rovibrational and rotational transitions
  within the X $^3\Sigma^-$ ground state of NH}, J. Chem. Phys. {141} ({2014})
  054310.
%\newblock \href {https://doi.org/{10.1063/1.4891468}} {\path{doi:{10.1063/1.4891468}}}.

\bibitem{15BrBeWe.NH}
J.~S.~A. Brooke, P.~F. Bernath, C.~M. Western, {Note: Improved line strengths
  of rovibrational and rotational transitions within the X~$^3\Sigma^-$ ground
  state of NH}, J. Chem. Phys. {143} ({2015}) 026101.
%\newblock \href {https://doi.org/{10.1063/1.4923422}} {\path{doi:{10.1063/1.4923422}}}.

\bibitem{18FeBeHo.NH}
A.~M. Fernando, P.~F. Bernath, J.~N. Hodges, T.~Masseron, {A new linelist for
  the A~$^3\Pi$ -- X~$^3\Sigma^-$ transition of the NH free radical}, J. Quant.
  Spectrosc. Radiat. Transf. {217} ({2018}) 29--34 .
%\newblock \href {https://doi.org/10.1016/j.jqsrt.2018.05.021} {\path{doi:10.1016/j.jqsrt.2018.05.021}}.

\bibitem{70MaBrGu}
J.~Malicet, J.~Brion, H.~Guenebau, Spectroscopic study of {A\,$^3\Pi$(I) --
  X\,$^3\sigma^-$S} transition in {NH} radicals, J. Chimie Physique
  Physico-Chimie Biologique {67} ({1970}) 24 .

\bibitem{75RaLi}
H.~E. Radford, M.~M. LitvakI, {Imine (NH) detected by laser magnetic
  resonance}, Chem. Phys. Lett. {34} ({1975}) 561--564.
%\newblock \href {https://doi.org/{10.1016/0009-2614(75)85562-X}} {\path{doi:{10.1016/0009-2614(75)85562-X}}}.

\bibitem{76WaRa}
F.~D. Wayne, H.~E. Radford, Laser magnetic-resonance spectra of imine ({NH})
  and its isotopes, Mol. Phys. {32} ({1976}) 1407--1422.
%\newblock \href {https://doi.org/{10.1080/00268977600102771}} {\path{doi:{10.1080/00268977600102771}}}.

\bibitem{78GrLe}
W.~R.~M. Graham, H.~Lew, {Spectra of the d\,$^1\Sigma^+$ --c\,$^1\Pi$ and
  d\,$^1\Sigma$--b\,$^1\Sigma^+$ systems and dissociation energy of NH and ND},
  Can. J. Phys. 56 (1978) 85--99.

\bibitem{82BeAm}
P.~F. Bernath, T.~Amano, {Difference frequency laser spectroscopy of the
  $v=1-0$ transition of NH}, J. Mol. Spectrosc. 95 (1982) 359--364.

\bibitem{82RaSa}
D.~A. Ramsay, P.~J. Sarre, {The c\,$^1\Pi$ -- a\,$^1\Delta$ system of the NH
  molecule}, J. Mol. Spectrosc. 93 (1982) 445--446.

\bibitem{82VaMeDy}
F.~C. Van~den Heuvel, W.~L. Meerts, A.~Dymanus, {Rotational hyperfine spectrum
  of the NH radical around 1 THz}, Chem. Phys. Lett. 92 (1982) 215--218.

\bibitem{84UbMeDy}
W.~Ubachs, J.~J. ter Meulen, A.~Dymanus, {High-resolution laser spectroscopy on
  the A\,$^3\Pi$ -- X\,$^3\Sigma^-$ transition of NH}, Can. J. Phys. 62 (1984)
  1374--1391.

\bibitem{85HaAdKaCu}
J.~L. Hall, H.~Adams, J.~V.~V. Kasper, R.~F. Curl, F.~K. Tittel, {Color-center
  laser kinetic spectroscopy: observation of the a\,$^1\Delta$ NH vibrational
  fundamental}, J. Opt. Soc. Am. B 2 (1985) 781--785.

\bibitem{86RaBe}
R.~S. Ram, P.~F. Bernath, {Fourier-transform spectroscopy of NH: the c $^1\Pi$
  -- a $^1\Delta$ transition}, J. Opt. Soc. Am. B 3 (1986) 1170--1174.

\bibitem{86BrRaBe}
C.~R. Brazier, R.~S. Ram, P.~F. Bernath, {Fourier transform spectroscopy of the
  A\,$^3\Pi$ -- X\,$^3\Sigma^-$ transition of NH}, J. Mol. Spectrosc. 120
  (1986) 381--402.
%\newblock \href {https://doi.org/10.1016/0022-2852(86)90012-3} {\path{doi:10.1016/0022-2852(86)90012-3}}.

\bibitem{86LeEvBr}
K.~R. Leopold, K.~M. Evenson, J.~M. Brown, {Far infrared laser magnetic
  resonance detection of NH and ND (a\,$^1\Delta$)}, J. Chem. Phys. 85 (1986)
  324--330.

\bibitem{86UbMeTeDy}
W.~Ubachs, G.~Meyer, J.~J. Ter~Meulen, A.~Dymanus, {High-resolution
  spectroscopy on the c\,$^1\Pi$ -- a\,$^1\Delta$ transition in NH}, J. Mol.
  Spectrosc. 115 (1986) 88--104.

\bibitem{86BoBrChGu}
D.~Boudjaadar, J.~Brion, P.~Chollet, G.~Guelachvilli, M.~Vervloet,
  Infrared-emission spectra of 5 {$\Delta v = 1$} sequence bands of the
  free-radical {NH} in its {X\,$^3\Sigma^-$} state, J. Mol. Spectrosc. {119}
  ({1986}) 352--366.
%\newblock \href {https://doi.org/{10.1016/0022-2852(86)90030-5}}  {\path{doi:{10.1016/0022-2852(86)90030-5}}}.

\bibitem{87VaDaBrEv}
E.~C.~C. Vasconcellos, S.~A. Davidson, J.~M. Brown, K.~R. Leopold, K.~M.
  Evenson, {Rotational and hyperfine constants of vibrationally excited
  NH(a\,$^1\Delta; v = 1$)}, J. Mol. Spectrosc. 122 (1987) 242 -- 245.
%\newblock \href {https://doi.org/10.1016/0022-2852(87)90233-5} {\path{doi:10.1016/0022-2852(87)90233-5}}.

\bibitem{90HaMi}
W.~Hack, T.~Mill, {Spectroscopic constants of NH(a\,$^1\Delta$) from the
  c\,$^1\Pi(v'= 0)$ - a\,$^1\Delta(v" \leq 3)$ laser-induced fluorescence
  spectra}, J. Mol. Spectrosc. 144 (1990) 358 -- 365.
%\newblock \href {https://doi.org/10.1016/0022-2852(90)90222-C} {\path{doi:10.1016/0022-2852(90)90222-C}}.

\bibitem{97KlTaWi}
T.~{Klaus}, S.~{Takano}, G.~{Winnewisser}, {Laboratory measurement of the
  $N=1-0$ rotational transition of NH at 1~THz.}, Astrophys. J. 322 (1997)
  L1--L4.

\bibitem{99RaBeHi}
R.~S. Ram, P.~F. Bernath, K.~H. Hinkle, {Infrared emission spectroscopy of NH:
  Comparison of a cryogenic echelle spectrograph with a Fourier transform
  spectrometer}, J. Chem. Phys. 110 (1999) 5557--5563.

\bibitem{99RiGe}
J.~L. Rinnenthal, K.-H. Gericke, {Direct high-resolution determination of the
  singlet--triplet splitting in NH using stimulated emission pumping}, J. Mol.
  Spectrosc. 198 (1999) 115--122.

\bibitem{03VaSaMoJo}
S.~Y.~T. van~de Meerakker, B.~G. Sartakov, A.~P. Mosk, R.~T. Jongma, G.~Meijer,
  {Optical pumping of metastable NH radicals into the paramagnetic ground
  state}, Phys. Rev. A 68 (2003) 032508.

\bibitem{04FlBrMaOd}
J.~Flores-Mijangos, J.~M. Brown, F.~Matsushima, H.~Odashima, K.~Takagi, L.~R.
  Zink, K.~M. Evenson, {The far-infrared spectrum of the $^{14}$NH radical in
  its X $^3\Sigma^-$ state}, J. Mol. Spectrosc. 225 (2004) 189--195.

\bibitem{04LeBrWiSi}
F.~Lewen, S.~Br\"unken, G.~Winnewisser, M.~{\v{S}}ime{\v{c}}kov{\'a},
  {\v{S}}.~Urban, {Doppler-limited rotational spectrum of the NH radical in the
  2 THz region}, J. Mol. Spectrosc. 226 (2004) 113--122.

\bibitem{07RoBrFlZi}
A.~Robinson, J.~Brown, J.~Flores-Mijangos, L.~Zink, M.~Jackson, {Spectroscopic
  study of the $^{14}$NH radical in vibrationally excited levels of the
  X~$^3\Sigma^-$ state by far infrared laser magnetic resonance}, Mol. Phys.
  105 (2007) 639--662.

\bibitem{10RaBe}
R.~S. Ram, P.~F. Bernath, {Revised molecular constants and term values for the
  X $^3\Sigma^-$ and A $^3\Pi$ states of NH}, J. Mol. Spectrosc. 260 (2010)
  115--119.

\bibitem{90GrLaSaVa}
N.~Grevesse, D.~L. Lambert, A.~J. Sauval, E.~F. {Van Dishoeck}, C.~B. Farmer,
  R.~H. Norton, Identification of solar vibration-rotation lines of {NH} and
  the solar nitrogen abundance, Astron. Astrophys. {232} ({1990}) 225--230.

\bibitem{91GeSaGrFa}
M.~{Geller}, C.~B. {Farmer}, R.~H. {Norton}, A.~J. {Sauval}, N.~{Grevesse},
  {First identification of pure rotation lines of NH in the infrared solar
  spectrum}, Astron. Astrophys. 249 (1991) 550--552.

\bibitem{jt412}
T.~Furtenbacher, A.~G. {Cs\'asz\'ar}, J.~Tennyson, {MARVEL: measured active
  rotational-vibrational energy levels}, J. Mol. Spectrosc. 245 (2007) 115--125
 .
%\newblock \href {https://doi.org/10.1016/j.jms.2007.07.005} {\path{doi:10.1016/j.jms.2007.07.005}}.

\bibitem{07CsCzFu.marvel}
A.~G. Cs{\'a}sz{\'a}r, G.~Czak{\'o}, T.~Furtenbacher, E.~M{\'a}tyus, An active
  database approach to complete rotational--vibrational spectra of small
  molecules, Annu. Rep. Comput. Chem. 3 (2007) 155--176.

\bibitem{12FuCsi.method}
T.~Furtenbacher, A.~G. {Cs\'asz\'ar}, {MARVEL: measured active
  rotational-vibrational energy levels. II. Algorithmic improvements}, J.
  Quant. Spectrosc. Radiat. Transf. 113 (2012) 929--935.

\bibitem{jt750}
R.~T\'obi\'as, T.~Furtenbacher, J.~Tennyson, A.~G. Cs\'asz\'ar, {Accurate
  empirical rovibrational energies and transitions of H$_2$$^{16}$O}, Phys.
  Chem. Chem. Phys. 21 (2019) 3473--3495.
%\newblock \href {https://doi.org/10.1039/c8cp05169k} {\path{doi:10.1039/c8cp05169k}}.

\bibitem{jt661}
T.~Furtenbacher, T.~Szidarovszky, J.~Hruby, A.~A. Kyuberis, N.~F. Zobov, O.~L.
  Polyansky, J.~Tennyson, A.~G. Cs\'asz\'ar, {Definitive high-temperature
  ideal-gas thermochemical functions of the H$_2$$^{16}$O molecule}, J. Phys.
  Chem. Ref. Data 45 (2016) 043104.
%\newblock \href {https://doi.org/10.1063/1.4967723} {\path{doi:10.1063/1.4967723}}.

\bibitem{jt743}
P.~A. Coles, R.~I. Ovsyannikov, O.~L. Polyansky, S.~N. Yurchenko, J.~Tennyson,
  {Improved potential energy surface and spectral assignments for ammonia in
  the near-infrared region}, J. Quant. Spectrosc. Radiat. Transf. 219 (2018)
  199--212.
%\newblock \href {https://doi.org/10.1016/j.jqsrt.2018.07.022} {\path{doi:10.1016/j.jqsrt.2018.07.022}}.

\bibitem{jt734}
O.~L. Polyansky, A.~A. Kyuberis, N.~F. Zobov, J.~Tennyson, S.~N. Yurchenko,
  L.~Lodi, {ExoMol molecular line lists XXX: a complete high-accuracy line list
  for water}, Mon. Not. R. Astron. Soc. 480 (2018) 2597--2608.
%\newblock \href {https://doi.org/10.1093/mnras/sty1877} {\path{doi:10.1093/mnras/sty1877}}.

\bibitem{19HuScLe.SO2}
X.~Huang, D.~W. Schwenke, T.~J. Lee, {Quantitative validation of Ames IR
  intensity and new line lists for $^{32/33/34}$S$^{16}$O$_2$,
  $^{32}$S$^{18}$O$_2$ and $^{16}$O$^{32}$S$^{18}$O}, J. Quant. Spectrosc.
  Radiat. Transf. 225 (2019) 327 -- 336.
%\newblock \href {https://doi.org/10.1016/j.jqsrt.2018.11.039} {\path{doi:10.1016/j.jqsrt.2018.11.039}}.

\bibitem{07OwJaKw}
L.~C. Owono~Owono, N.~Jaidane, M.~G. Kwato~Njock, Z.~Ben~Lakhdar, {Theoretical
  investigation of excited and Rydberg states of imidogen radical NH: Potential
  energy curves, spectroscopic constants, and dipole moment functions}, J.
  Chem. Phys. 126 (2007) 244302.

\bibitem{16SoShSu}
Z.~Song, D.~Shi, J.~Sun, Z.~Zhu, {Accurate spectroscopic calculations of the 12
  $\Lambda$-S and 25 $\Omega$ states of the NH radical including the spin-orbit
  coupling effect}, Comput. Theor. Chem. 1093 (2016) 81--90.

\bibitem{71NeSc}
S.~V. O'Neil, H.~F. Schaefer, {Configuration interaction study of the
  X~$^3\Sigma^-$, a~$^1\Delta$, and b~$^1\Sigma^+$ states of NH}, J. Chem.
  Phys. 55 (1971) 394--401.
%\newblock \href {https://doi.org/10.1063/1.1675534} {\path{doi:10.1063/1.1675534}}.

\bibitem{75MeRo}
W.~Meyer, P.~Rosmus, {PNO–CI and CEPA studies of electron correlation
  effects. III. Spectroscopic constants and dipole moment functions for the
  ground states of the first‐row and second‐row diatomic hydrides}, J.
  Chem. Phys. 63 (1975) 2356--2375.
%\newblock \href {https://doi.org/10.1063/1.431665} {\path{doi:10.1063/1.431665}}.

\bibitem{15Koput}
J.~Koput, Ab initio ground-state potential energy function and
  vibration-rotation energy levels of imidogen, {NH}, J. Comput. Chem. 36
  (2015) 1286--1294.

\bibitem{11CsFuxx.marvel}
A.~G. {Cs\'asz\'ar}, T.~Furtenbacher, Spectroscopic networks, J. Mol.
  Spectrosc. 266 (2011) 99 -- 103.
%\newblock \href {https://doi.org/10.1016/j.jms.2011.03.031} {\path{doi:10.1016/j.jms.2011.03.031}}.

\bibitem{12FuCsxx.marvel}
T.~Furtenbacher, A.~G. {Cs\'asz\'ar}, The role of intensities in determining
  characteristics of spectroscopic networks, J. Molec. Struct. 1009 (2012) 123
  -- 129.
%\newblock \href {https://doi.org/10.1016/j.molstruc.2011.10.057} {\path{doi:10.1016/j.molstruc.2011.10.057}}.

\bibitem{14FuArMe.marvel}
T.~Furtenbacher, P.~{\'A}rend{\'a}s, G.~Mellau, A.~G. Cs{\'a}sz{\'a}r, Simple
  molecules as complex systems, Sci. Rep. 4 (2014) 4654.

\bibitem{16ArPeFu.marvel}
P.~{\'A}rend{\'a}s, T.~Furtenbacher, A.~G. Cs{\'a}sz{\'a}r, On spectra of
  spectra, J. Math. Chem. 54 (2016) 806--822.
%\newblock \href {https://doi.org/10.1007/s10910-016-0591-1} {\path{doi:10.1007/s10910-016-0591-1}}.

\bibitem{jt672}
L.~K. McKemmish, T.~Masseron, S.~Sheppard, E.~Sandeman, Z.~Schofield,
  T.~Furtenbacher, A.~G. {Cs\'asz\'ar}, J.~Tennyson, C.~Sousa-Silva, {MARVEL
  analysis of the measured high-resolution spectra of $^{48}$Ti$^{16}$O},
  Astrophys. J. Suppl. 228 (2017) 15.
%\newblock \href {https://doi.org/10.3847/1538-4365/228/2/15} {\path{doi:10.3847/1538-4365/228/2/15}}.

\bibitem{jt760}
L.~K. McKemmish, T.~Masseron, J.~Hoeijmakers, V.~V. P\'{e}rez-Mesa, S.~L.
  Grimm, S.~N. Yurchenko, J.~Tennyson, {ExoMol molecular linelists -- XXXIII.
  The spectrum of titanium oxide}, Mon. Not. R. Astron. Soc. (2019).

\bibitem{jt637}
T.~Furtenbacher, I.~Szab{\'o}, A.~G. Cs{\'a}sz{\'a}r, P.~F. Bernath, S.~N.
  Yurchenko, J.~Tennyson, Experimental energy levels and partition function of
  the {$^{12}$C$_2$} molecule, Astrophys. J. Suppl. 224 (2016) 44.
%\newblock \href {https://doi.org/10.3847/0067-0049/224/2/44} {\path{doi:10.3847/0067-0049/224/2/44}}.

\bibitem{jt705}
K.~L. Chubb, M.~Joseph, J.~Franklin, N.~Choudhury, T.~Furtenbacher, A.~G.
  Cs\'asz\'ar, G.~Gaspard, P.~Oguoko, A.~Kelly, S.~N. Yurchenko, J.~Tennyson,
  C.~Sousa-Silva, {MARVEL analysis of the measured high-resolution spectra of
  C$_2$H$_2$}, J. Quant. Spectrosc. Radiat. Transf. 204 (2018) 42--55.
%\newblock \href {https://doi.org/10.1016/j.jqsrt.2017.08.018}  {\path{doi:10.1016/j.jqsrt.2017.08.018}}.

\bibitem{jt608}
A.~R. {Al Derzi}, T.~Furtenbacher, S.~N. Yurchenko, J.~Tennyson, A.~G.
  Cs\'asz\'ar, {MARVEL analysis of the measured high-resolution spectra of
  $^{14}$NH$_3$}, J. Quant. Spectrosc. Radiat. Transf. 161 (2015) 117--130.
%\newblock \href {https://doi.org/10.1016/j.jqsrt.2015.03.034} {\path{doi:10.1016/j.jqsrt.2015.03.034}}.

\bibitem{jtNH3update}
T.~Furtenbacher, P.~A. Coles, J.~Tennyson, A.~G. Cs\'asz\'ar, {Updated MARVEL
  energy levels for ammonia}, J. Quant. Spectrosc. Radiat. Transf.{To be
  submitted} (2019).

\bibitem{jt704}
R.~T\'obi\'as, T.~Furtenbacher, A.~G. Cs\'asz\'ar, O.~V. Naumenko, J.~Tennyson,
  J.-M. Flaud, P.~Kumard, B.~Poirier, {Critical evaluation of measured
  rotational-vibrational transitions of four sulphur isotopologues of
  S$^{16}$O$_2$}, J. Quant. Spectrosc. Radiat. Transf. 208 (2018) 152--163.
%\newblock \href {https://doi.org/10.1016/j.jqsrt.2018.01.006} {\path{doi:10.1016/j.jqsrt.2018.01.006}}.

\bibitem{jt718}
K.~L. Chubb, O.~V. Naumenko, S.~Keely, S.~Bartolotto, S.~MacDonald, M.~Mukhtar,
  A.~Grachov, J.~White, E.~Coleman, S.-M. Hu, A.~Liu, A.~Z. Fazliev, E.~R.
  Polovtseva, V.~M. Horneman, A.~Campargue, T.~Furtenbacher, A.~G. Cs\'asz\'ar,
  S.~N. Yurchenko, J.~Tennyson, {MARVEL analysis of the measured
  high-resolution rovibrational spectra of H$_2$S}, J. Quant. Spectrosc.
  Radiat. Transf. 218 (2018) 178 -- 186.
%\newblock \href {https://doi.org/10.1016/j.jqsrt.2018.07.012} {\path{doi:10.1016/j.jqsrt.2018.07.012}}.

\bibitem{jt740}
L.~K. McKemmish, J.~Borsovszky, K.~L. Goodhew, S.~Sheppard, A.~F.~V. Bennett,
  A.~D.~J. Martin, A.~Singh, C.~A.~J. Sturgeon, T.~Furtenbacher, A.~G.
  Cs\'asz\'ar, J.~Tennyson, {MARVEL analysis of the measured high-resolution
  spectra of $^{90}$Zr$^{16}$O}, Astrophys. J. Suppl. 867 (2018) 33.
%\newblock \href {https://doi.org/10.3847/1538-4357/aadd19} {\path{doi:10.3847/1538-4357/aadd19}}.

\bibitem{13FuSzMa.marvel}
T.~Furtenbacher, T.~Szidarovszky, E.~M{\'a}tyus, C.~F{\'a}bri, A.~G.
  Cs{\'a}sz{\'a}r, {Analysis of the rotational--vibrational states of the
  molecular ion H$_3^+$}, J. Chem. Theory Comput. 9 (2013) 5471--5478.
%\newblock \href {https://doi.org/10.1021/ct4004355} {\path{doi:10.1021/ct4004355}}.

\bibitem{13FuSzFa.marvel}
T.~Furtenbacher, T.~Szidarovszky, C.~F{\'a}bri, A.~G. Cs{\'a}sz{\'a}r, {MARVEL
  analysis of the rotational--vibrational states of the molecular ions
  H$_2$D$^+$ and D$_2$H$^+$}, Phys. Chem. Chem. Phys. 15 (2013) 10181--10193.
%\newblock \href {https://doi.org/10.1039/c3cp44610g} {\path{doi:10.1039/c3cp44610g}}.

\bibitem{jt454}
J.~Tennyson, P.~F. Bernath, L.~R. Brown, A.~Campargue, M.~R. Carleer, A.~G.
  Cs\'asz\'ar, R.~R. Gamache, J.~T. Hodges, A.~Jenouvrier, O.~V. Naumenko,
  O.~L. Polyansky, L.~S. Rothman, R.~A. Toth, A.~C. Vandaele, N.~F. Zobov,
  L.~Daumont, A.~Z. Fazliev, T.~Furtenbacher, I.~E. Gordon, S.~N. Mikhailenko,
  S.~V. Shirin, {IUPAC critical evaluation of the rotational-vibrational
  spectra of water vapor. Part I. Energy levels and transition wavenumbers for
  H$_2$$^{17}$O and H$_2$$^{18}$O}, J. Quant. Spectrosc. Radiat. Transf. 110
  (2009) 573--596.
%\newblock \href {https://doi.org/10.1016/j.jqsrt.2009.02.014} {\path{doi:10.1016/j.jqsrt.2009.02.014}}.

\bibitem{jt482}
J.~Tennyson, P.~F. Bernath, L.~R. Brown, A.~Campargue, M.~R. Carleer, A.~G.
  Cs\'asz\'ar, L.~Daumont, R.~R. Gamache, J.~T. Hodges, O.~V. Naumenko, O.~L.
  Polyansky, L.~S. Rothman, R.~A. Toth, A.~C. Vandaele, N.~F. Zobov, A.~Z.
  Fazliev, T.~Furtenbacher, I.~E. Gordon, S.~N. Mikhailenko, B.~A. Voronin,
  {IUPAC critical evaluation of the rotational-vibrational spectra of water
  vapor. Part II. Energy levels and transition wavenumbers for HD$^{16}$O,
  HD$^{17}$O, and HD$^{18}$O}, J. Quant. Spectrosc. Radiat. Transf. 111 (2010)
  2160--2184.
%\newblock \href {https://doi.org/10.1016/j.jqsrt.2010.06.012} {\path{doi:10.1016/j.jqsrt.2010.06.012}}.

\bibitem{jt539}
J.~Tennyson, P.~F. Bernath, L.~R. Brown, A.~Campargue, M.~R. Carleer, A.~G.
  Cs\'asz\'ar, L.~Daumont, R.~R. Gamache, J.~T. Hodges, O.~V. Naumenko, O.~L.
  Polyansky, L.~S. Rothmam, A.~C. Vandaele, N.~F. Zobov, A.~R. {Al Derzi},
  C.~F\'abri, A.~Z. Fazliev, T.~Furtenbacher, I.~E. Gordon, L.~Lodi, I.~I.
  Mizus, {IUPAC critical evaluation of the rotational-vibrational spectra of
  water vapor. Part III. Energy levels and transition wavenumbers for
  H$_2$$^{16}$O}, J. Quant. Spectrosc. Radiat. Transf. 117 (2013) 29--80.
%\newblock \href {https://doi.org/10.1016/j.jqsrt.2012.10.002} {\path{doi:10.1016/j.jqsrt.2012.10.002}}.

\bibitem{jt576}
J.~Tennyson, P.~F. Bernath, L.~R. Brown, A.~Campargue, A.~G. Cs\'asz\'ar,
  L.~Daumont, R.~R. Gamache, J.~T. Hodges, O.~V. Naumenko, O.~L. Polyansky,
  L.~S. Rothmam, A.~C. Vandaele, N.~F. Zobov, N.~D\'enes, A.~Z. Fazliev,
  T.~Furtenbacher, I.~E. Gordon, S.-M. Hu, T.~Szidarovszky, I.~A. Vasilenko,
  {IUPAC critical evaluation of the rotational-vibrational spectra of water
  vapor. Part IV. Energy levels and transition wavenumbers for D$_2$$^{16}$O,
  D$_2$$^{17}$O and D$_2$$^{18}$O}, J. Quant. Spectrosc. Radiat. Transf. 142
  (2014) 93--108.
%\newblock \href {https://doi.org/10.1016/j.jqsrt.2014.03.019} {\path{doi:10.1016/j.jqsrt.2014.03.019}}.

\bibitem{jt562}
J.~Tennyson, P.~F. Bernath, L.~R. Brown, A.~Campargue, A.~G. Cs\'asz\'ar,
  L.~Daumont, R.~R. Gamache, J.~T. Hodges, O.~V. Naumenko, O.~L. Polyansky,
  L.~S. Rothman, A.~C. Vandaele, N.~F. Zobov, {A database of water transitions
  from experiment and theory (IUPAC Technical Report)}, Pure Appl. Chem. 86
  (2014) 71--83.
%\newblock \href {https://doi.org/10.1515/pac-2014-5012} {\path{doi:10.1515/pac-2014-5012}}.

\bibitem{jtwaterupdate}
T.~Furtenbacher, J.~Tennyson, O.~V. Naumenko, O.~L. Polyansky, N.~F. Zobov,
  A.~G. Cs\'asz\'ar, {The 2018 update of the IUPAC database of water energy
  levels}, J. Quant. Spectrosc. Radiat. Transf.(In preparation) (2019).

\bibitem{75BrHoHu.diatom}
J.~M. Brown, J.~T. Hougen, K.~P. Huber, J.~W.~C. Johns, I.~Kopp,
  H.~Lefebvre-Brion, A.~J. Merer, D.~A. Ramsay, J.~Rostas, R.~N. Zare, Labeling
  of parity doublet levels in linear molecules, J. Mol. Spectrosc. 55 (1975)
  500--503.
%\newblock \href {https://doi.org/10.1016/0022-2852(75)90291-X} {\path{doi:10.1016/0022-2852(75)90291-X}}.

\bibitem{96RaBe}
R.~S. Ram, P.~F. Bernath, {Fourier transform infrared emission spectroscopy of
  ND and PH}, J. Mol. Spectrosc. 176 (1996) 329--336.

\bibitem{59Dixon}
R.~N. Dixon, The $0-0$ and $1-0$ bands of the {A\,$^3\Pi_i$ -- X\,$^3\Sigma^-$}
  system of {NH}, Can. J. Phys. {37} ({1959}) 1171.

\bibitem{17ToFuCs}
R.~T\'obi\'as, T.~Furtenbacher, A.~G. Cs{\'a}sz{\'a}r, Cycle basis to the
  rescue, J. Quant. Spectrosc. Rad. Transfer 203 (2017) 557--564.

\bibitem{69KrNa}
G.~Krishnamurty, N.~A. Narasimham, {Predissociations in the d $^1\Sigma^-$ - c
  $^1\Pi$ bands of NH}, J. Mol. Spectrosc. 29 (1969) 410--414.

\bibitem{ATMOS}
M.~Abrams, A.~Goldman, M.~Gunson, C.~Rinsland, R.~Zander, {Observations of the
  infrared solar spectrum from space by the ATMOS experiment}, Appl. Optics
  {35} ({1996}) 2747--2751.
%\newblock \href {https://doi.org/{10.1364/AO.35.002747}} {\path{doi:{10.1364/AO.35.002747}}}.

\bibitem{ACE2}
P.~F. Bernath, {The Atmospheric Chemistry Experiment (ACE)}, J. Quant.
  Spectrosc. Radiat. Transf. {186} ({2017}) 3--16.
%\newblock \href {https://doi.org/{10.1016/j.jqsrt.2016.04.006}} {\path{doi:{10.1016/j.jqsrt.2016.04.006}}}.

\bibitem{10HaWaMc}
F.~Hase, L.~Wallace, S.~D. McLeod, J.~J. Harrison, P.~F. Bernath, {The ACE-FTS
  atlas of the infrared solar spectrum}, J. Quant. Spectrosc. Radiat. Transf.
  {111} ({2010}) 521--528.
%\newblock \href {https://doi.org/{10.1016/j.jqsrt.2009.10.020}} {\path{doi:{10.1016/j.jqsrt.2009.10.020}}}.

\bibitem{84RoSt}
F.~Rohrer, F.~Stuhl, {NH (a~$^1\Delta$ -- X~$^3\Sigma^-$) emission from the
  gas-phase photolysis of HN$_3$}, Chem. Phys. Lett. 111 (1984) 234--237.

\bibitem{PGOPHER}
C.~M. Western, {PGOPHER}: A program for simulating rotational, vibrational and
  electronic spectra, J. Quant. Spectrosc. Radiat. Transf. 186 (2017) 221--242.
%\newblock \href {https://doi.org/10.1016/j.jqsrt.2016.04.010}  {\path{doi:10.1016/j.jqsrt.2016.04.010}}.

\bibitem{74GiMaVe}
A.~Gilles, J.~Masanet, C.~Vermeil, {Direct determination of the NH
  b~$^1\Sigma^+$ -- X~$^3\Sigma^-$ energy difference}, Chem. Phys. Lett. 25
  (1974) 346--347.

\bibitem{63McHeEhGo}
B.~McBride, \href{https://books.google.hu/books?id=o3WGHAAACAAJ}{Thermodynamic
  Properties to 6000 degrees K for 210 substances involving the first 18
  elements}, National Aeronautics and Space Administration special paper,
  Office of Scientific and Technical Information, National Aeronautics and
  Space Administration (1963).
\newline\urlprefix\url{https://books.google.hu/books?id=o3WGHAAACAAJ}

\bibitem{79HuHe}
K.~Huber, G.~Herzberg, Molecular Spectra and Molecular Structure. IV. Constants
  of Diatomic Molecule, Van Nostrand Reinhold Company, New York (1979).

\bibitem{Gurvich2}
L.~V. Gurvich, I.~V. Veyts, C.~B. {Alcock (eds.)}, Thermodynamic Properties of
  Individual Substances, {V}ol. 1, {P}art two: {T}ables, Hemisphere Publishing
  Corporation, New York (1989).

\bibitem{93AlEaCs}
W.~D. Allen, A.~L.~L. East, A.~G. Cs\'asz\'ar, Ab initio anharmonic vibrational analyses of non-rigid molecules,
in: J.~Laane, M.~Dakkouri,
  B.~van~der Veken, H.~Oberhammer (Eds.), Structures and conformations of
  nonrigid molecules, Kluwer, Dordrecht, 1993, pp. 343--373.

\bibitem{98CsAlSc}
A.~G. {Cs\'asz\'ar}, W.~D. Allen, H.~F. {Schaefer III}, In pursuit of the ab
  initio limit for conformational energy prototypes, J. Chem. Phys. 108 (1998)
  9751--9764.

\bibitem{91AsClHo}
M.~N.~R. Ashfold, S.~G. Clement, J.~D. Howe, C.~M. Western, {Resonance-enhanced
  multiphoton ionisation spectroscopy of the NH (ND) radical. Part 1. The d
  $^1\Sigma^+$ state}, J. Chem. Soc. Faraday Trans. 87 (1991) 2515--2523.

\bibitem{92ClAsWe}
S.~G. Clement, M.~N.~R. Ashfold, C.~M. Western, {Resonance-enhanced multiphoton
  ionisation spectroscopy of the NH (ND) radical. Part 2.—Singlet members of
  the 3p Rydberg complex}, J. Chem. Soc. Faraday Trans. 88~(21) (1992)
  3121--3128.

\bibitem{92ClAsWeJo}
S.~G. Clement, M.~N.~R. Ashfold, C.~M. Western, R.~D. Johnson~III, J.~W.
  Hudgens, Triplet rydberg states of the imidogen radical characterized via
  two-photon resonance-enhanced multiphoton ionization spectroscopy, J. Chem.
  Phys. 97 (1992) 7064--7072.

\bibitem{92ClAsWeDe}
S.~G. Clement, M.~N.~R. Ashfold, C.~M. Western, E.~de~Beer, C.~A. de~Lange,
  N.~P.~C. Westwood, {New singlet Rydberg states of the NH (ND) radical in the
  energy range 92 000--100 000 cm$^{-1}$ characterized by resonance enhanced
  multiphoton ionization-photoelectron spectroscopy}, J. Chem. Phys. 96 (1992)
  4963--4973.

\bibitem{jtMARVEL}
T.~Furtenbacher, A.~G. Cs\'asz\'ar, J.~Tennyson, et~al., {The MARVEL project},
  J. Quant. Spectrosc. Radiat. Transf.{To be submitted} (2019).

\bibitem{jtCH4Marvel}
E.~J. Barton, M.~Liu, T.~Farnell, J.~Tennyson, C.~Sousa-Silva, V.~Boudon,
  T.~Furtenbacher, A.~G. Cs{\'a}sz{\'a}r, {MARVEL analysis of the measured
  rotation-vibration spectra of methane}, JQSRT (2019).

\bibitem{jt709}
C.~Sousa-Silva, L.~K. McKemmish, K.~L. Chubb, J.~Baker, E.~J. Barton, M.~N.
  Gorman, T.~Rivlin, J.~Tennyson, {Original Research By Young Twinkle Students
  (ORBYTS): When can students start performing original research?}, Phys. Educ.
  53 (2018) 015020.
%\newblock \href {https://doi.org/10.1088/1361-6552/aa8f2a} {\path{doi:10.1088/1361-6552/aa8f2a}}.

\bibitem{19FuHoKoSo}
T.~Furtenbacher, M.~Horv\'ath, D.~Koller, P.~S\'olyom, A.~Balogh, I.~Balogh,
  A.~G. {Cs\'asz\'ar}, {MARVEL} analysis of the measured high-resolution
  rovibronic spectra and definitive ideal-gas thermochemistry of the
  {$^{16}$O$_2$} molecule, J. Phys. Chem. Ref. Data 48 (2019) 023101.
%\newblock \href {https://doi.org/10.1063/1.5083135} {\path{doi:10.1063/1.5083135}}.

\end{thebibliography}

\end{document}